\documentclass[%
aps,
superscriptaddress,
 amsmath,amssymb,
 reprint,%
]{revtex4-1}

\usepackage{graphicx}
\usepackage{dcolumn}
\usepackage{bm}
\usepackage[english]{babel}
\usepackage[utf8]{inputenc}
\usepackage[T1]{fontenc}
\usepackage{xcolor}
\usepackage{comment}
\usepackage{mathtools}
\usepackage{amsmath}
\usepackage{mathrsfs}
\usepackage{microtype}

\DeclareFontFamily{U}{calligra}{}
\DeclareFontShape{U}{calligra}{m}{n}{<->callig15}{}

\begin{document}

\title{Broken intrinsic symmetry induced magnon-magnon coupling in synthetic ferrimagnets}

\author{Mohammad Tomal Hossain}
\affiliation{Department of Physics and Astronomy, University of Delaware, Newark, Delaware 19716, USA}

\author{Hang Chen}
\affiliation{Department of Physics and Astronomy, University of Delaware, Newark, Delaware 19716, USA}

\author{Subhash Bhatt}
\affiliation{Department of Physics and Astronomy, University of Delaware, Newark, Delaware 19716, USA}

\author{Mojtaba Taghipour Kaffash}
\affiliation{Department of Physics and Astronomy, University of Delaware, Newark, Delaware 19716, USA}

\author{Mitra M. Subedi}
\affiliation{Department of Physics and Astronomy, Wayne State University, Detroit, MI 48202, USA}

\author{John Q. Xiao}
\affiliation{Department of Physics and Astronomy, University of Delaware, Newark, Delaware 19716, USA}

\author{Joseph Sklenar}
\affiliation{Department of Physics and Astronomy, Wayne State University, Detroit, MI 48202, USA}

\author{M. Benjamin Jungfleisch}
\email[]{mbj@udel.edu}
\affiliation{Department of Physics and Astronomy, University of Delaware, Newark, Delaware 19716, USA}

\date{\today}

\begin{abstract}

Synthetic antiferromagnets offer rich magnon energy spectra in which optical and acoustic magnon branches can hybridize.  Here, we demonstrate a broken intrinsic symmetry induced coupling of acoustic and optical magnons in a synthetic ferrimagnet consisting of two dissimilar antiferromagnetically interacting ferromagnetic metals. Two distinct magnon modes hybridize at degeneracy points, as indicated by an avoided level-crossing. The avoided level-crossing gap depends on the interlayer exchange interaction between the magnetic layers, which can be controlled by adjusting the non-magnetic interlayer thickness. A large avoided level crossing gap of 3.9 GHz is revealed, exceeding the coupling strength that is typically found in other magnonic hybrid systems based on a coupling of magnons with photons or magnons with phonons. 

\end{abstract}

\maketitle

\section{Introduction}
Magnons are the quanta of spin waves representing the fundamental excitation of the magnetic order in a magnetic system. Magnons play a crucial role in our understanding of magnetism. They have been the subject of extensive study for their properties and applications in various technologies, particularly in the development of novel magnonic devices \cite{chen_unidirectional_2021, haldar_reconfigurable_2016, urazhdin_nanomagnonic_2014, khitun_magnonic_2010, mahmoud_introduction_2020, chumak_advances_2022, wang_nanoscale_2024}. 
Antiferromagnetic magnons are investigated for their potential use in ultrafast applications including next-generation memory devices and information processing \cite{xiong_antiferromagnetic_2022, kosub_purely_2017, olejnik_antiferromagnetic_2017, olejnik_terahertz_2018,jungfleisch_high-frequency_2017}, terahertz spectroscopy and imaging \cite{lee_terahertz_2021, rongione_emission_2023, mashkovich_terahertz_2019, bowlan_using_2018,wu_principles_2021}, and quantum computing \cite{meier_quantum_2003}. 
Giant magnetoresistance (GMR) was discovered in 1988 in antiferromagnetically coupled layered ferromagnetic (FM) structures \cite{baibich_giant_1988,binasch_enhanced_1989}. This marks one of the initial applications of synthetic antiferromagnetic (sAF) heterostructures comprising two ferromagnetic metal layers coupled antiferromagnetically. This idea expanded to various other devices such as magnetic tunnel junctions and spin-torque nano-oscillators \cite{ grollier_spintronic_2016, kiselev_microwave_2003, luo_implementation_2023}.
sAFs offer several engineering advantages compared to bulk antiferromagnets. Unlike bulk antiferromagnetic materials, sAFs can be tailored with specific layer thicknesses and compositions, which allows for the optimization of magnetic properties such as exchange bias, coercivity, and anisotropy. These engineering opportunities make sAFs ideal model systems for studying antiferromagnetic magnons.

Coupling between magnon modes in sAFs can lead to the exchange of energy and momentum between different magnon modes. By controlling the coupling strength, the propagation and transmission of spin waves can be manipulated. This can improve performance in magnonic devices such as magnon transistor \cite{gong_electrically_2018}, and magnon logic elements \cite{chumak_fundamentals_2019}. 
The most dominant mechanisms responsible for magnon-magnon coupling \cite{zhang_perspective_2025} are the exchange interaction (short range) \cite{macneill_gigahertz_2019} and the dipolar interaction (long range) \cite{shiota_tunable_2020}. In sAFs comprising two sublattices with antiferromagnetic exchange interaction, two distinct normal modes can emerge based on the relative motion of the spins precessing in the sublattices; in-phase motion (acoustic mode) or out-of-phase motion (optical mode). Generally, these modes are protected by symmetry, preventing them from coupling \cite{neumann_uber_1929, patchett_symmetry_2022}. In a symmetric synthetic antiferromagnet, where the two ferromagnetic layers are magnetically equivalent, a $180^\circ$ rotation about the field direction (when the field is aligned with a high-symmetry axis) is a valid symmetry operation. Under this operation, the acoustic mode is odd ($A\rightarrow -A$) and the optical mode is even ($A\rightarrow A$). Any coupling term must be invariant under the symmetry operation; however, the product of an even and an odd mode changes sign and therefore the coupling vanishes, allowing the modes to cross without hybridization. Strategies to facilitate magnon-magnon coupling in sAFs have been studied recently \cite{sud_tunable_2020, dai_strong_2021,hayashi_observation_2023, shiota_tunable_2020,sklenar_self-hybridization_2021,chen_tuning_2020}.

Furthermore, a recent theoretical work suggested the possibility of magnon-magnon coupling for an in-plane geometry in an intrinsic symmetry-breaking mechanism using a synthetic ferrimagnetic structure \cite{li_symmetry_2021}. Ferrimagnets are partially compensated magnetic structures in which the sublattices with unequal magnetic moments are arranged in opposite directions, which results in non-zero spontaneous magnetization. In a synthetic structure, this can be achieved in different ways, such as by using FM layers with different saturation magnetization, $M_\mathrm{s}$, by using different thicknesses of the FM layers, $d$, by using an unequal number of layers, or any combination of them, all connected by negative Ruderman–Kittel–Kasuya–Yosida (RKKY) \footnote{also known as bilinear or quadratic interaction} interaction between layers \cite{lepadatu_synthetic_2017, xie_engineering_2023, jenkins_current_2014, franco_multi-state_2018, sud_magnon-magnon_2023,wang_ultrastrong_2024,subedi_even-odd-layer-dependent_2024}. Synthetic ferrimagnets can facilitate intrinsic symmetry breaking in the presence of RKKY and biquadratic interactions, which leads to the coupling of acoustic and optical modes \cite{li_symmetry_2021}.

\begin{figure}[b]
    \centering
    \includegraphics[width=0.99\columnwidth]{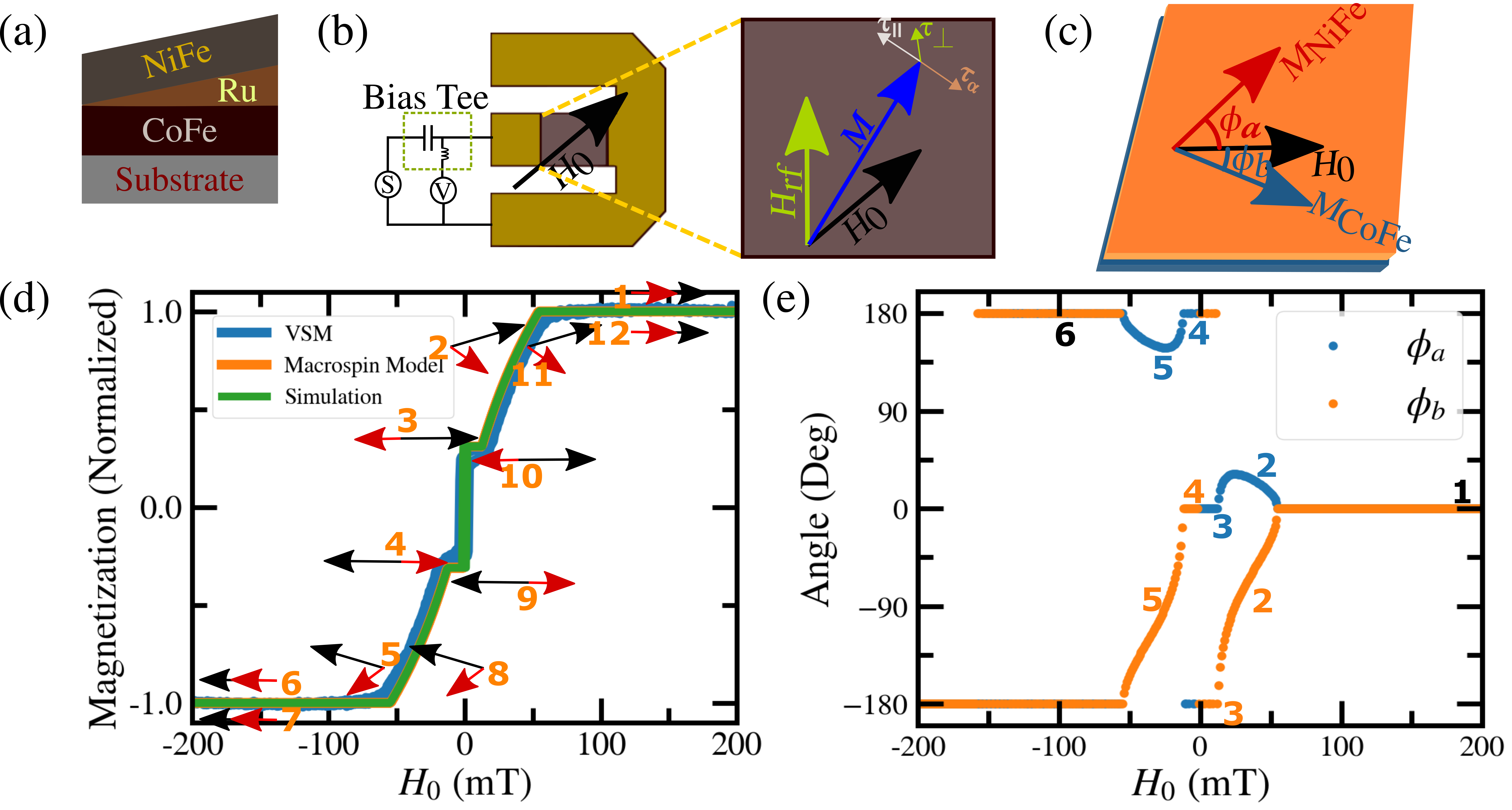}
        \caption{(a) Schematic illustration of the synthetic ferrimagnet comprising of a CoFe/Ru/NiFe heterostructure. The thickness of Ru is varied from $0$~nm to $1$~nm to tune the coupling strength between CoFe and NiFe. (b) Illustration of the spin-torque ferromagnetic resonance (STFMR) setup, including an inset that shows the relevant torques on the magnetization $\boldsymbol{M}$: $\boldsymbol{\tau}_{||}$ - anti-damping-like torque, $\boldsymbol{\tau}_\alpha$ - damping-like torque, $\boldsymbol{\tau}_\perp$ - field-like torque. The biasing field $\boldsymbol{H}_0$ is applied at an angle of 45 degrees with respect to the microwave field $\boldsymbol{H}_\mathrm{rf}$. (c) Schematic diagram of equilibrium angles of the ferromagnetic layers relative to the external field direction. (d) Vibrating sample magnetometry (VSM) measurement for a sample with Ru thickness of $0.8$~nm. Corresponding macrospin model calculations and micromagnetic simulations (MuMax3) are overlaid. Both the macrospin model and simulations show good agreement with experiment. The arrows indicate the relative alignment of the two exchange-coupled ferromagnetic layers. (e) MuMax3 simulations of the equilibrium angles of the ferromagnetic layers as a function of the applied field during field ramp-down, referenced to the external field direction defined in (c). Note that -180° and 180° correspond to the same magnetization direction. In some regions, the blue data points (magnetization direction of CoFe) are obscured by the overlapping orange data points (magnetization direction of NiFe).
    }
    \label{fig:fig1}
\end{figure}

Here, we report the experimental observation of such magnon-magnon coupling induced by broken intrinsic symmetry in synthetic ferrimagnetic structures. We use two antiferromagnetically interacting magnetic layers of similar thickness, $d$, but different saturation magnetization, $M_\mathrm{s}$ (CoFe and NiFe) separated by a thin spacer layer (Ru). Magnetometry using vibrating sample magnetometry (VSM) reveals the exchange interaction between the ferromagnetic layers. Spin dynamics spectrum reveals that the asymmetry introduced by broken intrinsic symmetry facilitates the coupling between the acoustic and optical magnon modes. Furthermore, our studies show that the coupling strength between the modes can be tuned by controlling the exchange interaction between the ferromagnetic layers. Our experimental results are systematically compared with macrospin modeling and micromagnetic simulations using MuMax3 \cite{vansteenkiste_design_2014}. Our study aims to leverage the magnon-magnon coupling within an in-plane geometry to develop innovative magnonic device applications.

\section{Sample Fabrication and Experimental Details}
Figure~\ref{fig:fig1}(a) illustrates a schematic of the synthetic ferrimagnetic sample stack. The samples comprise a trilayer of $\mathrm{Co_{90}}\mathrm{Fe_{10}(10~nm)}$/$\mathrm{Ru(}d~\mathrm{ nm)}$/$\mathrm{Ni_{80}}\mathrm{Fe_{20}(10~nm)}$ with varying thickness $d$ of the non-magnetic spacer layer Ru. The structure was grown on thermally oxidized Si substrates at room temperature under an Ar pressure of $\mathrm{4.5\times{10^{-3}}}$ Torr in a magnetron sputtering system with a base pressure of $\mathrm{1.4\times{10^{-7}}}$ Torr. The Ru spacer layer was grown in a wedge shape, allowing for a continuous thickness variation from $d=0$ to 1 nm at an wedge angle of about $\mathrm{1~}\mathrm{\mu} \mathrm{deg}$, by off-center deposition to tune the interlayer exchange interaction. All samples were capped with a 5 nm $\mathrm{SiO}_\mathrm{2}$ layer to protect the sample from oxidation. The sample was then cut into 11 rectangular samples where the thickness of the spacer layer can be considered approximately constant over the width of each sample. The anisotropic easy axes in the FM layers were introduced along the long sides of the samples. 
VSM measurements were performed to characterize the static magnetic properties of the individual pieces of the sample. Figure~\ref{fig:fig1}(d) shows a typical VSM measurement of a synthetic ferrimagnet with negative RKKY interaction between the ferromagnetic layers, which we will discuss below.

To probe the spin dynamics of the ferrimagnet, we employ spin-torque ferromagnetic resonance (STFMR) spectroscopy. The STFMR signal is the spin rectification of the magnetoresistance in the device \cite{gui_realization_2007}, which provides high sensitivity measurements.
In STFMR measurements a spin-polarized current in addition to an RF magnetic field exerts a torque on the magnetization, causing it to precess when the resonance condition is realized. This precession induces a time-varying magnetization, consequently giving rise to an AC resistance as a result of the anisotropic magnetoresistance or spin Hall magnetoresistance \cite{tulapurkar_spin-torque_2005, sankey_spin-transfer-driven_2006}. This, combined with the AC current passed through the sample at the same frequency, generates a rectified DC voltage, which is measured using a lock-in amplifier \cite{liu_spin-torque_2011}.
For STFMR measurements, the sputtered thin films were etched to produce STFMR devices following the following steps: first, a negative-tone photoresist was used to cover 80 $\mathrm{\mu}$m $\times$ 130~$\mathrm{\mu}$m rectangular sections of the film, followed by etching using Ar ion milling of the surrounding film. The shorted coplanar waveguides were then patterned on top of the rectangular sections using optical lithography and positive photoresist. Finally, 5~nm of Ti and 20~nm of Au were e-beam evaporated and lifted off to create the waveguides on the devices as shown in Fig.~\ref{fig:fig1}(b).

\begin{figure}[t]
    \centering
    \includegraphics[width=0.9\columnwidth]{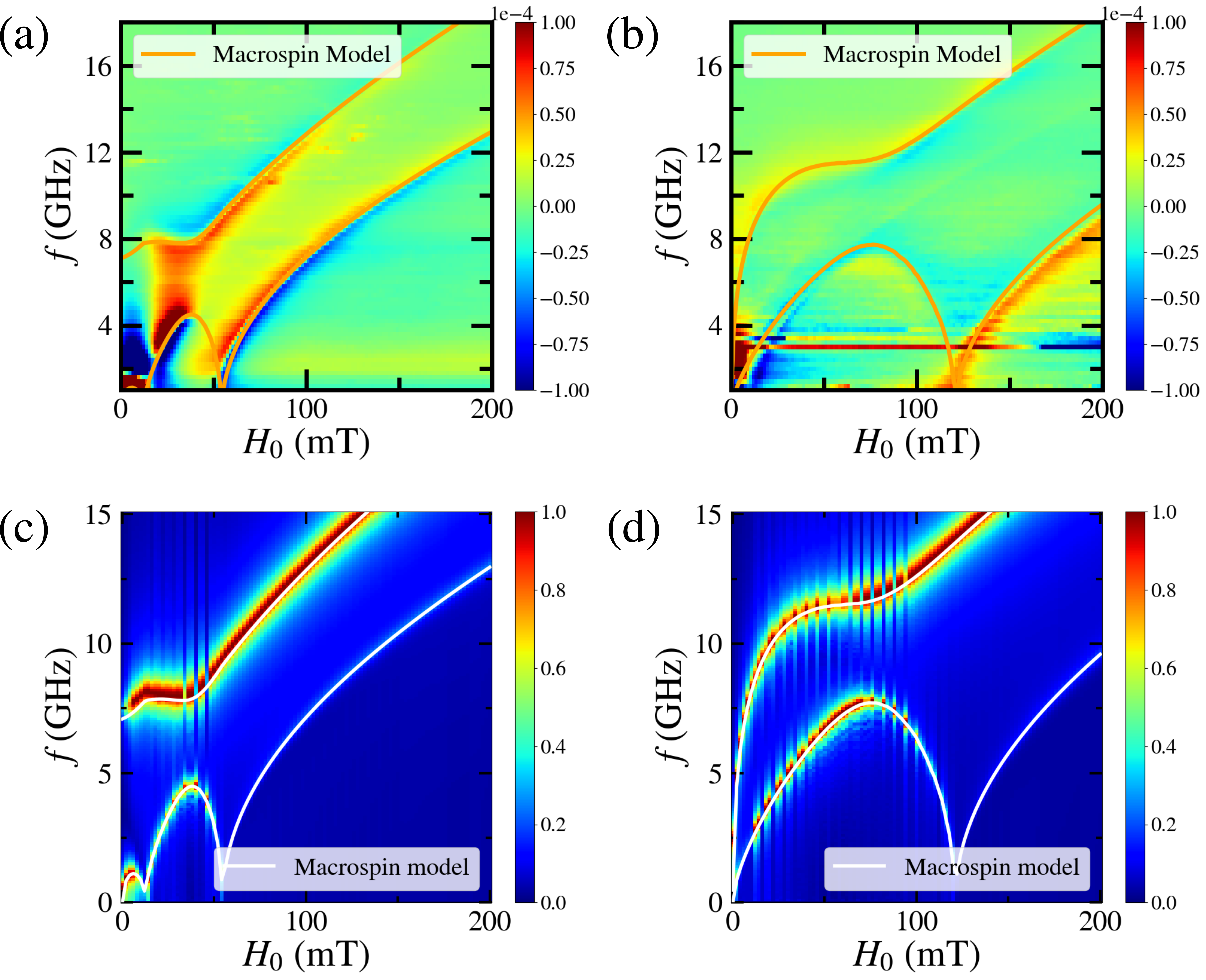}
    \caption{(a,b) Experimentally measured magnon spectra for Ru thicknesses of 0.8~nm and 0.4~nm, with corresponding macrospin model calculations overlaid. The color scale represents the lock-in amplifier output. (c,d) Corresponding micromagnetic (MuMax3) simulations of the STFMR spectra, with macrospin model results shown for comparison. Here, the color scale represents the FFT amplitude.
}
    \label{fig:fig2}
\end{figure}

A bias tee was employed to simultaneously apply a microwave current and measure the rectified DC voltage using a lock-in amplifier~\cite{hossain_probing_2021}. A microwave signal with an output power of +21~dBm was generated by an Agilent E8257D signal generator and amplitude-modulated at a frequency of 666.7~Hz. The external magnetic field, \( \boldsymbol{H}_\mathrm{0} \), was applied in the plane of the sample at an angle of \( 45^{\circ} \) relative to the device axis [see Fig.~\ref{fig:fig1}(b)] to maximize the signal amplitude~\cite{harder_analysis_2011, liu_spin-torque_2011}. For each applied field, the magnetic moments of the two ferromagnetic layers (NiFe and CoFe) reach equilibrium orientations characterized by angles \(\phi_\mathrm{a}\) and \(\phi_\mathrm{b}\), respectively, with respect to the external field direction \(\boldsymbol{H}_0\) [see Fig.~\ref{fig:fig1}(c)]. These equilibrium angles were determined from micromagnetic simulations [see Fig.~\ref{fig:fig1}(e)]. The microwave frequency was swept from 1~GHz to 18~GHz in 86 steps, while the external magnetic field was varied from +200~mT to 0~mT for each frequency step.

\begin{figure*}[t]
    \centering
    \includegraphics[width=0.8\paperwidth]{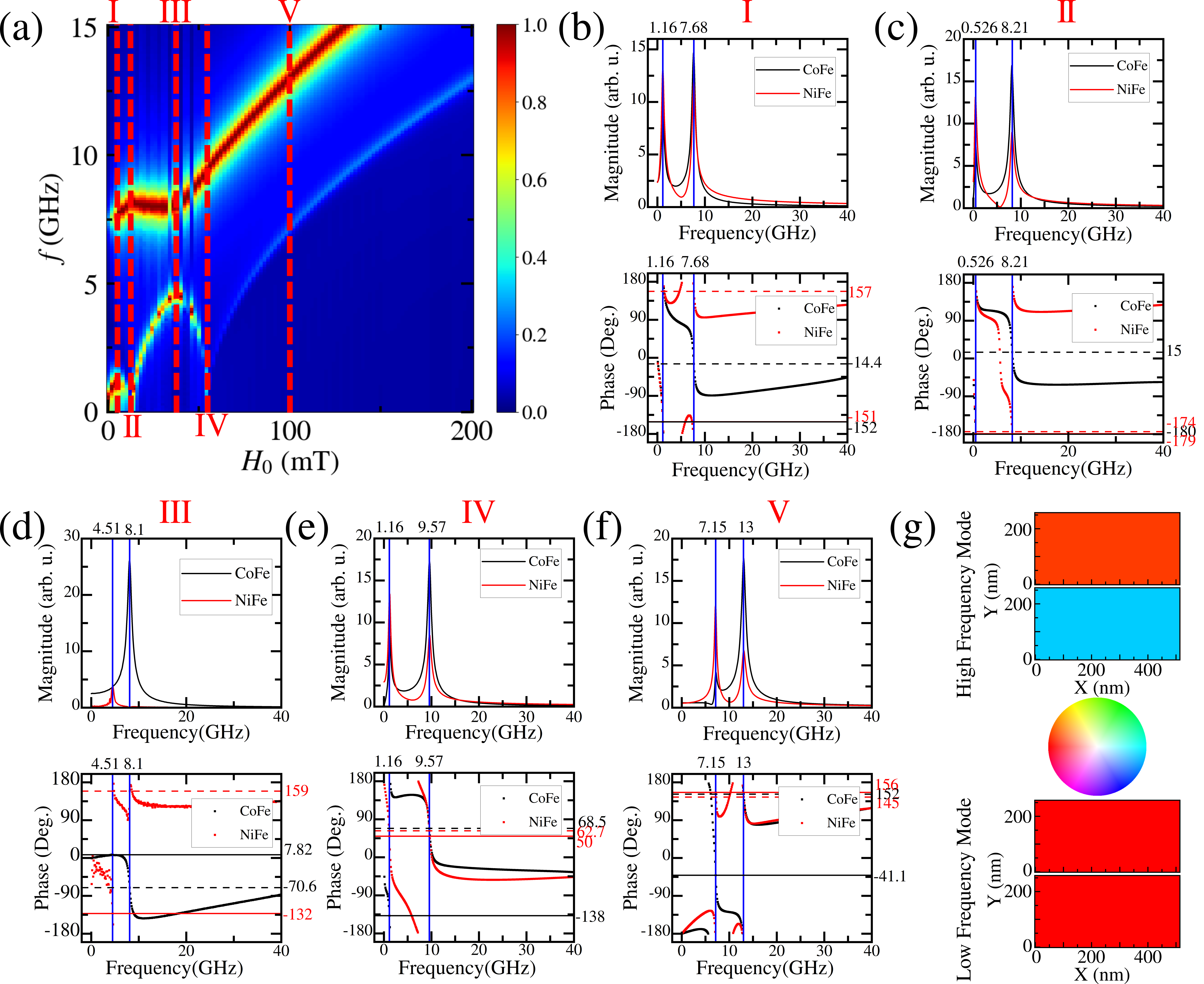}
    \caption{Micromagnetic simulations obtained with MuMax3. (a) Simulated spectrum for a Ru thickness of 0.8~nm. (b–f) Layer-resolved simulations showing the amplitude (top) and phase (bottom) profiles of the magnon modes, extracted from FFTs of the magnetization dynamics at 4, 12, 37, 56, and 100~mT. The two vertical lines in each panel indicate the resonance frequencies at that field. These lines intersect the phase spectra of CoFe (black) and NiFe (red) at the lower-frequency (solid) and higher-frequency (dashed) branches, highlighted by the corresponding horizontal markers in the phase panels. (g) Spatially resolved simulation at 12~mT with X and Y indicating the lateral dimensions of the studied devices. The color wheel represents the magnetization dynamics phase.
    }
    \label{fig:fig3}
\end{figure*}

\section{Results and Discussion}
As is shown in Fig. \ref{fig:fig1}(d), magnetometry measurements show a typical hysteresis loop of a synthetic antiferromagnet. Magnetometry measurements show secondary loops in the hysteresis for Ru thicknesses above $0.2$~nm [see Fig. S2 in the supplementary materials (SM)]. From $0.2-0.5$~nm, two secondary loops appear near the \textit{knee} (the field at which the magnetization changes from a saturated to an unsaturated state), while above 0.5~nm, a loop forms at the center [Fig. \ref{fig:fig1}(d)]. The `\textit{knee} loops' suggest the presence of strong biquadratic exchange interaction between FM films for Ru thicknesses of $0.2-0.5$~nm \cite{demokritov_biquadratic_1998} (see SM). The magnetometry results can be explained by carefully analyzing the expression of the volume energy density of the structure \cite{inoue_theory_1994}:

\begin{equation}
    \mathrm{\mathscr{E}}=-\boldsymbol{\mathrm{M}} \cdot \boldsymbol{\mathrm{H_0}}+J_\mathrm{q} \boldsymbol{m}_1\cdot\boldsymbol{m}_2+J_\mathrm{{bq}}(\boldsymbol{m}_1\cdot\boldsymbol{m}_2)^2.
    \label{eq.1}
\end{equation}

Here, $\boldsymbol{M}$ is the magnetization, $\boldsymbol{H}_\mathrm{0}$ is the external magnetic field, and $\boldsymbol{m}_\mathrm{i}$ are the magnetization unit vectors of the $i$-th layer ($i=1,2$), respectively.  $J_\mathrm{q}$ and $J_\mathrm{{bq}}$ are the phenomenological energy densities for quadratic and biquadratic exchange interaction. RKKY interaction is negative (i.e., favoring antiferromagnetic alignment of the magnetic layers) when $J_\mathrm{q}$ is positive. At a large field, the magnetostatic energy is large enough to saturate both layers in the direction of the field [configuration labeled `1' in Fig.~\ref{fig:fig1}(d)]. As the field is reduced, a field is reached [beginning of configuration `2' in Fig.~\ref{fig:fig1}(d)] where the exchange energy becomes significant and attempts to align the FM layers antiferromagnetically. As a result, the two layers reach an equilibrium at a tilted angle with respect to each other away from the saturated state, reducing the net magnetization. The critical field at which the magnetization starts to reduce (i.e., the \textit{knee} field in the broader hysteresis loop) can be calculated using a macrospin model as in Eq.~(\ref{eq.1}) (see SM for the derivation):
    \begin{equation}
        \mathrm{H_\mathrm{{0,ff}}=(J_\mathrm{q}+2J_\mathrm{{bq}})\left(\frac{1}{M_\mathrm{{S1}}}+\frac{1}{M_\mathrm{{S2}}}\right)},
        \tag{2a}
        \label{eq.2a}
    \end{equation}
where $H_\mathrm{{0,ff}}$ is the magnitude of the magnetic field over which the FM layers align in a forced ferromagnetic state [configuration `1' in Fig.~\ref{fig:fig1}(d)], and $M_\mathrm{Si}$ are the saturation magnetizations of the $i$-th layer ($i=1,2$), respectively.

If the biquadratic exchange interaction is negligible, magnetization changes linearly with the external field in the unsaturated configuration [configuration `2' in Fig.~\ref{fig:fig1}(d)] (see SM for derivation). However, strong biquadratic interaction causes a non-linear relation, creating secondary `knee loops' near the knee field in the hysteresis (see SM).

As the external field decreases further, the quadratic exchange energy dominates, forming a plateau [configuration `3' in Fig.~\ref{fig:fig1}(d)] with antiparallel layer alignment. The finite magnetization at this plateau results from different saturation magnetizations between the ferromagnetic layers \cite{waring_zero-field_2020}, creating the `central loop' in the magnetometry data, as seen in Fig.~\ref{fig:fig1}(d). The critical field at which the layers align antiferromagnetically (i.e., the \textit{plateau} region in the central part of the hysteresis loop) is given by Eq.~(\ref{eq.2b}):
\begin{equation}
    H_\mathrm{0,afm} = (J_\mathrm{q} - 2J_\mathrm{bq})\left(\frac{1}{M_\mathrm{S2}} - \frac{1}{M_\mathrm{S1}}\right),
    \tag{2b}
    \label{eq.2b}
\end{equation}
where, when the external magnetic field lies between \(-H_\mathrm{0,afm}\) and \(H_\mathrm{0,afm}\), the ferromagnetic layers align antiferromagnetically [configuration `3' in Fig.~\ref{fig:fig1}(d)], and \( M_\mathrm{S1} > M_\mathrm{S2} \). Equation~(\ref{eq.2b}) explains the absence of a central loop in magnetometry when Ru thickness is less than $0.6$~nm, given $J_\mathrm{q}<2J_\mathrm{bq}$ makes $H_\mathrm{0,afm}$ negative, preventing a fully antiferromagnetic state. Eqs.~(\ref{eq.2a}) and (\ref{eq.2b}) provide a means to extract $J_\mathrm{q}$ and $J_\mathrm{bq}$ from magnetometry measurements when $J_\mathrm{q}>2J_\mathrm{bq}$, allowing a fully antiferromagnetic state:

   \begin{equation}
       \mathrm{J_q=\left(\frac{H_{0,ff}}{M_{S1}+M_{S2}}+\frac{H_{0,afm}}{M_{S1}-M_{S2}}\right)\frac{M_{S1}M_{S2}}{2}},
       \tag{3a}
       \label{eq.3a}
   \end{equation}
   \begin{equation}
       \mathrm{J_{bq}=\left(\frac{H_{0,ff}}{M_{S1}+M_{S2}}-\frac{H_{0,afm}}{M_{S1}-M_{S2}}\right)\frac{M_{S1}M_{S2}}{4}}.
       \tag{3b}
       \label{eq.3b}
   \end{equation}

The relationship between the saturation magnetizations of the ferromagnetic layers can be established using the normalized magnetization of the forced ferromagnetic ($M_\mathrm{ff}$) and fully antiferromagnetic ($M_\mathrm{afm}$) states (the magnetizations at configuration `1' and configuration `3' in Fig.~\ref{fig:fig1}(d) respectively). The ratio of the saturation magnetizations can be expressed as follows:
    \begin{equation}
       M_\mathrm{r} = \mathrm{\frac{M_{S2}}{M_{S1}}=\frac{M_{ff}-M_{afm}}{M_{ff}+M_{afm}}}.
       \tag{4}
       \label{Ms}
    \end{equation}
For a Ru thickness of $0.8\,\mathrm{nm}$, we extracted the \textit{critical} fields from the VSM results [Fig.~\ref{fig:fig1}(d)] as
$H_{0,\mathrm{ff}} = 55 \pm 1.5\,\mathrm{mT}$ and 
$H_{0,\mathrm{afm}} = 13 \pm 1\,\mathrm{mT}$, 
with corresponding normalized magnetizations 
$M_{\mathrm{ff}} = 1.0$ and 
$M_{\mathrm{afm}} = 0.302 \pm 0.028$. 
From these values, the ratio of the saturation magnetization is calculated as 
$M_{\mathrm{r}} = 0.536 \pm 0.033$.

Figures~\ref{fig:fig2}(a,b) show the STFMR spectra for samples with Ru thicknesses of 0.8~nm and 0.4~nm, both exhibiting antiferromagnetic exchange interaction as confirmed by magnetometry (see Fig.~S2). We notice two magnon modes in the spectra of all of our samples with negative RKKY interaction (as found in magnetometry measurements; see also SM). These two modes correspond to the optical and acoustic modes (more details on the phase of the modes can be found below). Typically, these two modes are protected by symmetry, which means that they cross at the degeneracy point. As is evident from Fig.~\ref{fig:fig2}, that is not what we observe in the experiments: a clear avoided-level crossing is observed at the degeneracy points, indicating strong coupling between the optical and acoustic modes. The coupling gap is determined by fitting the resonance peaks in the STFMR spectra with standard Lorentzian functions (see SM for details) and identifying the minimum frequency separation between the resonances at a fixed magnetic field. The largest gap of 3.9 GHz is observed for the sample with a Ru thickness of 0.4 nm. The enhanced signal amplitude at 3.8 GHz is a measurement artifact and does not carry physical significance.

We model the magnetization dynamics by treating the two ferromagnetic layers as coupled macrospins, whose time evolution is governed by the Landau–Lifshitz–Gilbert (LLG) equation \cite{landau_theory_1935, gilbert_phenomenological_2004, stancil_spin_2008}:
\begin{align}
    \frac{d\mathbf{m}_\mathrm{1}}{dt} &= -\mu_0 \gamma \mathbf{m}_\mathrm{1} \times \Big[ 
        H_0 \hat{\mathbf{y}}
        - H_\mathrm{E1} \mathbf{m}_\mathrm{2} \notag \\
        &\qquad
        - H_\mathrm{B1} (\mathbf{m}_\mathrm{1} \cdot \mathbf{m}_\mathrm{2}) \mathbf{m}_\mathrm{2}
        - M_\mathrm{S1} (\mathbf{m}_\mathrm{1} \cdot \hat{\mathbf{z}}) \hat{\mathbf{z}}
    \Big].
    \tag{5a}\label{LLG1}
\end{align}
\begin{align}
    \frac{d\mathbf{m}_\mathrm{2}}{dt} &= -\mu_0 \gamma \mathbf{m}_\mathrm{2} \times \Big[
        H_0 \hat{\mathbf{y}}
        - H_\mathrm{E2} \mathbf{m}_\mathrm{1}
         \notag \\
        &\qquad
        - H_\mathrm{B2} (\mathbf{m}_\mathrm{1} \cdot \mathbf{m}_\mathrm{2}) \mathbf{m}_\mathrm{1} - M_\mathrm{S2} (\mathbf{m}_\mathrm{2} \cdot \hat{\mathbf{z}}) \hat{\mathbf{z}}
    \Big].
    \tag{5b}\label{LLG2}
\end{align}

Here, $\mathbf{m}_{1}$ and $\mathbf{m}_{2}$ denote the macrospin unit vectors of the ferromagnetic layers, with their dynamics described by 
\[
\frac{d\mathbf{m}_\mathrm{i}}{dt} = i \omega \, \delta \mathbf{m}_\mathrm{i},
\]
where \(\delta \mathbf{m}_\mathrm{i}\) is the unit vector of the dynamic deviation of the magnetization unit vector. The effective fields $H_\mathrm{E\,i}$ and $H_\mathrm{B\,i}$ due to quadratic and biquadratic exchange are defined as 
\[
H_\mathrm{E\,i} = \frac{J_\mathrm{q}}{M_\mathrm{S\,i}}, \qquad 
H_\mathrm{B\,i} = \frac{2J_\mathrm{bq}}{M_\mathrm{S\,i}}.
\]
Finally, the torque terms proportional to the saturation magnetizations $M_\mathrm{S1}$ and $M_\mathrm{S2}$ in Eqs.~\ref{LLG1} and \ref{LLG2} parameterize the shape-dependent
demagnetizing fields of the NiFe and CoFe layers, respectively.

After some algebra (see SM), the nontrivial solutions are obtained by requiring that the determinant of the coefficient matrix vanishes:
\begin{widetext}
    \begin{equation}
    \resizebox{0.99\textwidth}{!}{$
        \begin{vmatrix}
        i\omega & \mu_0 \gamma (M_1 + H_\mathrm{1,eq}) & 0 & \frac{\mu_0 \gamma}{M_1}( J_\mathrm{q} + 2J_\mathrm{bq} \cos{\alpha}) \\
        -\mu_0 \gamma \left(\frac{2J_\mathrm{bq}}{M_1} \sin^2{\alpha} + H_\mathrm{1,eq}\right) & i\omega & \frac{\mu_0 \gamma}{M_1}\left[2J_\mathrm{bq}( \sin^2{\alpha} -\cos^2{\alpha})- J_\mathrm{q} \cos{\alpha}\right]& 0\\
        0 & \frac{\mu_0 \gamma}{M_2}( J_\mathrm{q} + 2J_\mathrm{bq} \cos{\alpha})  & i\omega & \mu_0 \gamma (M_2+H_\mathrm{2,eq})\\
        \frac{\mu_0 \gamma}{M_2}\left[2J_\mathrm{bq}( \sin^2{\alpha} -\cos^2{\alpha})- J_\mathrm{q} \cos{\alpha}\right]& 0 & -\mu_0 \gamma \left(\frac{2J_\mathrm{bq}}{M_2} \sin^2{\alpha} + H_\mathrm{2,eq}\right) & i\omega\\
        \end{vmatrix}
        = 0,
    $}
    \tag{6}
    \label{Eigenvalue_eq}
    \end{equation}
\end{widetext}
where \(H_{\mathrm{i,eq}}\) denotes the effective magnetic field experienced by layer \(i\), and \(\alpha=\phi_\mathrm{a}+\phi_\mathrm{b}\) is the equilibrium interlayer angle.

Numerical solutions of the determinant yield the magnon branches for a given set of parameters, and the experimental spectra can be fitted within this framework to extract the corresponding material parameters. Fitting the STFMR spectra with the macrospin model is complicated by the hidden variables \(\phi_{\mathrm{a}}\) and \(\phi_{\mathrm{b}}\), which describe the equilibrium magnetization orientations of the two ferromagnetic layers and cannot be measured directly. Experimentally, we obtain the resonance frequency as a function of the external magnetic field, while the angles \(\phi_{\mathrm{a}}\) and \(\phi_{\mathrm{b}}\) must be determined self-consistently from the model. The fitting algorithm and corresponding Python code are provided in the SM.

 For a Ru thickness of $0.8\,\mathrm{nm}$, the fitted parameters are $M^{\mathrm{fit}}_{\mathrm{S1}} = 1.82 \times 10^{6}~\mathrm{A/m}$, $M^{\mathrm{fit}}_{\mathrm{r}} = 0.527$, $J^{\mathrm{fit}}_\mathrm{q} = 29.8~\mathrm{kJ/m^3}$, and 
$J^{\mathrm{fit}}_\mathrm{bq} = 2.16~\mathrm{kJ/m^3}$. The fitted macrospin magnon spectrum is overlaid with the STFMR data in Fig.~\ref{fig:fig2}(a). An analogous procedure was followed for the sample with a Ru thickness of \(0.4~\mathrm{nm}\), shown in Fig.~\ref{fig:fig2}(b), demonstrating good agreement in both cases. Using the material parameters obtained from the dynamic macrospin model, we compare the calculated hysteresis loop with the experimental VSM data shown in Fig.~\ref{fig:fig1}(d) and find good overall agreement. Finally, these parameters also yield the \textit{knee} fields from Eqs.~(\ref{eq.2a}) and (\ref{eq.2b}), resulting in \( H^{\mathrm{fit}}_{0,\mathrm{ff}} = 54.4~\mathrm{mT} \) and \( H^{\mathrm{fit}}_{0,\mathrm{afm}} = 12.6~\mathrm{mT} \), which are consistent with the values extracted from the VSM measurements ($H_{0,\mathrm{ff}} = 55 \pm 1.5\,\mathrm{mT}$ and 
$H_{0,\mathrm{afm}} = 13 \pm 1\,\mathrm{mT}$) within experimental uncertainty.

To gain a better understanding of the experimental results, we performed micromagnetic modeling using the graphics processor unit (GPU)-accelerated program MuMax3 \cite{vansteenkiste_design_2014, exl_labontes_2014}. Our structure was modeled as $128 \times 64 \times 3$ cells with an individual cell size of $4.0$~nm~$\times~ 4.0$~nm~$ \times~4.0$~nm, with the middle layer representing the spacer. Each layer was defined into a region to set material parameters separately. The values for the exchange stiffness of CoFe and NiFe used in the simulation were $3.0\times 10^{-11}$ J/m and $1.3\times 10^{-11}$ J/m, respectively \cite{parreiras_effect_2015}. This corresponds to exchange lengths for NiFe and CoFe of 5.29~nm and 4.32~nm,  respectively \cite{parreiras_effect_2015}. Crystalline anisotropy was considered negligible. The following Gilbert damping constants were used for CoFe and NiFe, respectively:
$\mathrm{\alpha_\mathrm{CoFe} = 0.005}$ \cite{weber_gilbert_2019} and $\mathrm{\alpha_\mathrm{NiFe} = 0.0095}$ \cite{hrabec_spin-orbit_2016}. We define the quadratic and biquadratic interlayer exchange interactions using custom expressions for the exchange field and the corresponding energy density. These expressions are constructed to reproduce the exchange energy given in Eq.~(\ref{eq.1}) (the third and fourth terms on the right-hand side) and to ensure that the resulting magnetization dynamics are consistent with the LLG equation. We define the quadratic and biquadratic exchange fields as (see SM): 
\begin{equation}
    \mathrm{\boldsymbol{\mathrm{H}}_\mathrm{quad}=-\frac{J_\mathrm{q}}{M_\mathrm{S1}M_\mathrm{S2}}\widetilde{\boldsymbol{\mathrm{M}}}},
    \tag{7a}
    \label{eq.5a}
    \end{equation}
\begin{equation}
   \mathrm{\boldsymbol{\mathrm{H}}_\mathrm{biquad}=-\frac{2J_\mathrm{bq}}{(M_\mathrm{S1}M_\mathrm{S2})^2}(\widetilde{\boldsymbol{\mathrm{M}}}\cdot\boldsymbol{\mathrm{M}})\widetilde{\boldsymbol{\mathrm{M}}}}.
   \tag{7b}
   \label{eq.5b}
\end{equation}

and corresponding energy density terms as:
\begin{equation}
    \mathrm{\mathscr{E}_\mathrm{quad}=- \mathrm{\boldsymbol{\mathrm{M}}}}\cdot\mathrm{\boldsymbol{\mathrm{H}}}_\mathrm{quad}.
    \tag{8a}
    \label{eq.8a}
\end{equation}
\begin{equation}
    \mathrm{\mathscr{E}_\mathrm{biquad}=- \frac{1}{2}\mathrm{\boldsymbol{\mathrm{M}}}}\cdot\mathrm{\boldsymbol{\mathrm{H}}}_\mathrm{biquad}.
    \tag{8b}
    \label{eq.8b}
\end{equation}

Here, $\widetilde{\boldsymbol{M}}$ is the swapped magnetization defined as $\widetilde{\boldsymbol{M}}(x,y,z,t)=\boldsymbol{M}(x,y,-z,t)$. For the simulations, the saturation magnetizations and exchange coupling constants were taken from macrospin model fits to the STFMR spectra.

We first performed magnetostatic simulations to generate the hysteresis loop. We started the simulation with a static magnetic field of $+200$~mT along the long axis, swept down to $-200$~mT in steps of $2$~mT, and ramped up back to $+200$~mT to complete the hysteresis loop. The built-in relaxation function in MuMax3 was used to find the equilibrium state of the system. The simulated hysteresis loops show good agreement with both the VSM measurements and the macrospin model, as illustrated in Fig.~\ref{fig:fig1}(d) for a Ru thickness of $0.8\,\mathrm{nm}$. MuMax3 simulations of the equilibrium angles of the ferromagnetic layers with respect to the external field, $\phi_\mathrm{a}$ (CoFe) and $\phi_\mathrm{b}$ (NiFe), are plotted in Fig.~\ref{fig:fig1}(e) as a function of the applied field during field ramp-down. Note that -180° and 180° correspond to the same magnetization direction. In some regions, the blue data points (magnetization direction of CoFe) are obscured by the overlapping orange data points (magnetization direction of NiFe). This plot provides direct insight into the transitions observed in the magnetostatic measurements and calculations shown in Fig.~\ref{fig:fig1}(d). In particular, the transition from configuration `1' to configuration `2' corresponds to a spin-flop transition, whereas the transition from configuration `3' to configuration `4' corresponds to a spin-flip transition.

\begin{figure}[t]
    \centering
    \includegraphics[width=0.95\columnwidth]{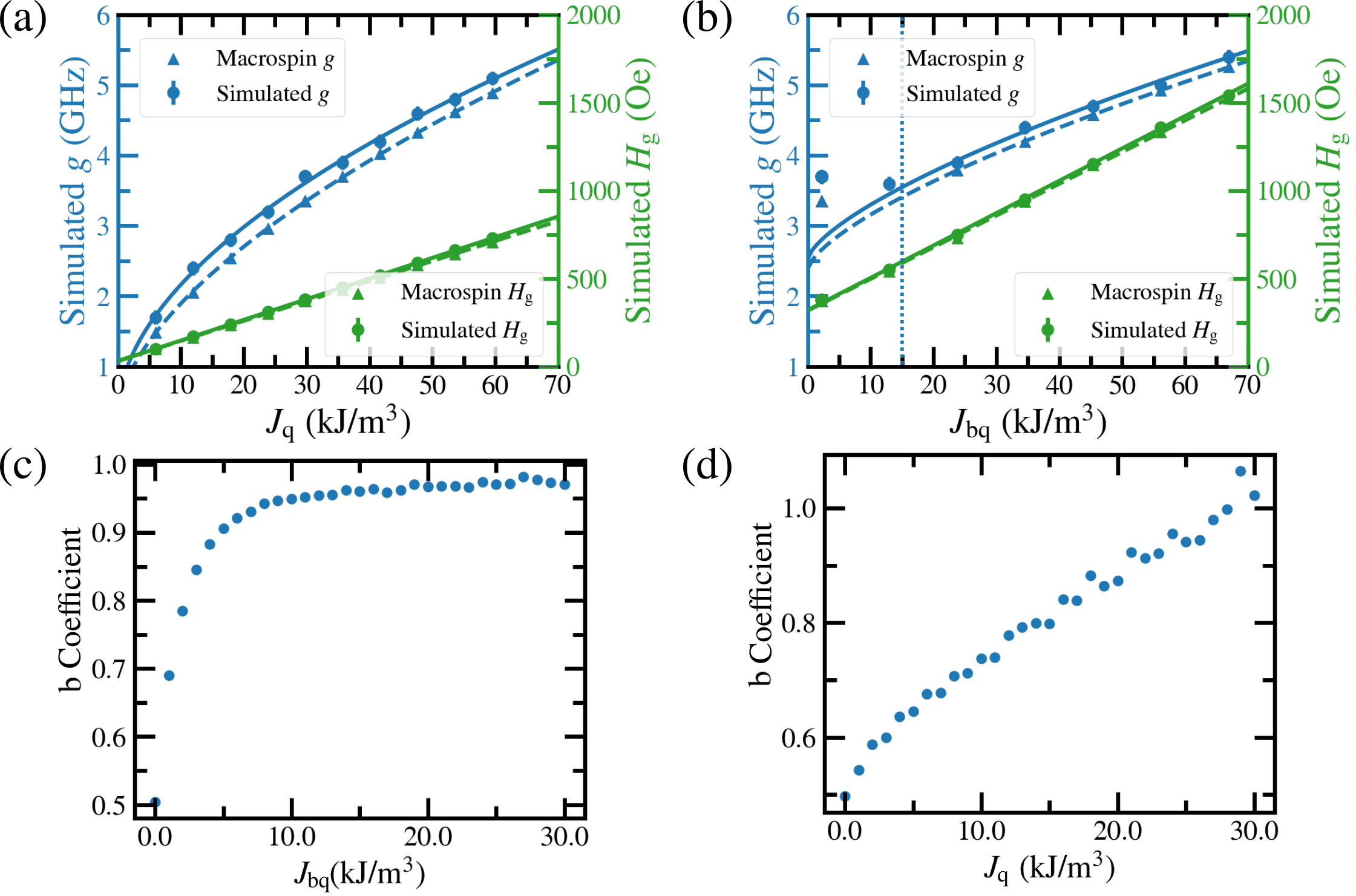}
    \caption{(a) Coupling gap \( g \) (blue) and degeneracy field \( H_\mathrm{g} \) (green) obtained from macrospin calculations (solid lines) and micromagnetic simulations (dashed lines) as a function of the quadratic exchange constant \( J_\mathrm{q} \) at a fixed biquadratic exchange constant of \( J_\mathrm{bq} = 2.16~\mathrm{kJ/m^3} \). The degeneracy field \( H_\mathrm{g} \) is defined as the magnetic field at which the frequency separation between the high- and low-frequency modes is minimized, yielding the coupling gap \( g \). The uncertainty (error bar) in the simulated \(H_g\) arises from the finite field resolution, given by the field step size (2~mT in this case). The uncertainty in the simulated \(g\) is determined by the finite frequency resolution of the FFT, set by the simulation runtime (100~MHz for 10~ns), and is multiplied by \(\sqrt{2}\) to account for error propagation. (b) Same as (a), but shown as a function of the biquadratic exchange constant \( J_\mathrm{bq} \) at a fixed quadratic exchange constant of \( J_\mathrm{q} = 29.8~\mathrm{kJ/m^3} \). (c--d) The b coefficient in the sublinear power-law dependence on \( J_\mathrm{bq} \) and \( J_\mathrm{q} \), respectively.
    }
    \label{fig:fig4}
\end{figure}

Dynamic simulations using MuMax3 reveal the phase relations between the ferromagnetic layers in the dynamic spectra. Figures~\ref{fig:fig2}(c,d) show simulated spectra for Ru thicknesses of 0.8~nm and 0.4~nm, corresponding to the experimental STFMR measurements in Figs.~\ref{fig:fig2}(a,b). We started the simulation with a static magnetic field of +200~mT and swept down to 0~mT in steps of 1~mT, with the initial magnetization state at each field taken from magnetostatic simulations. A sinc pulse with an amplitude of \(0.1~\mathrm{mT}\) and a cutoff frequency of 25~GHz was applied along all three spatial axes to simultaneously excite both the acoustic and optical modes~\cite{subedi_engineering_2025}, while a static external field was applied along the long axis of the sample. The RF field direction and phase were chosen to minimize numerical artifacts. At each field step, the dynamics were evolved for 10~ns, and the total magnetization of the coupled layers was Fourier transformed to obtain the spectra. Macrospin model calculations are overlaid on the simulations and show excellent agreement. The spectra show two hybridized magnon branches similar to the experimentally obtained spectra, one branch appearing more prominent than the other, consistent with the general observation that the acoustic mode is readily excited whereas the optical mode is more difficult to excite, particularly in the saturated regime~\cite{xiao_magnon_2022}. Importantly, the RF field direction and phase do not affect the spectra (see SM).

We further analyzed the FFT phase to determine the magnon mode relations at different fields. At 4~mT [configuration `3' in Fig.~\ref{fig:fig1}(d,e); see Fig.~\ref{fig:fig3}(b)], when the layers are antiferromagnetically aligned, the FFT magnitude (top panel) exhibits two peaks at 1.16~GHz and 7.68~GHz in both ferromagnetic layers, marked by vertical lines. At the lower frequency branch (1.16~GHz), the CoFe and NiFe phases (bottom panel) are $-151.6^\circ$ and $-151.3^\circ$, marked by solid horizontal lines, differing by only $0.3^\circ$, consistent with an in-phase acoustic mode. At the high-frequency branch (7.68~GHz), the phases are $-14.4^\circ$ (CoFe) and $157.3^\circ$ (NiFe), marked by dashed horizontal lines, differing by $171.7^\circ$, characteristic of an out-of-phase optical mode. 

At the transition field $H_\mathrm{0,afm}=12~\mathrm{mT}$ [transition point from configuration `2' to configuration `3' in Fig.~\ref{fig:fig1}(d,e); Fig.~\ref{fig:fig3}(c)], the phase differences remain close to $1^\circ$ (acoustic) and $189^\circ$ (optical). Near the degeneracy point, the FFT phase exhibits a rapid variation. For this reason, we performed simulations with higher magnetic-field resolution, using a step size of 0.2~mT. At 37.4~mT [Fig.~\ref{fig:fig3}(d)], we find an avoided level crossing and a gap in the spectrum, where the phase differences increase to $40^\circ$ and $230^\circ$, respectively, neither close to what we would expect from an acoustic or optical mode. The modes are no longer purely acoustic or optical in nature, indicating hybridization of the modes. At the second transition field $H_\mathrm{0,ff}=52~\mathrm{mT}$ [transition point from configuration `1' to configuration `2'  in Fig.~\ref{fig:fig1}(d,e); Fig.~\ref{fig:fig3}(e)], the phase differences switch to $188^\circ$ (low frequency) and $6^\circ$ (high frequency), corresponding to an interchange of acoustic and optical character of the two modes. In the saturated regime [configuration `1'; Fig.~\ref{fig:fig3}(f)], at 100~mT the phase differences are $197^\circ$ and $7^\circ$, again indicating optical and acoustic modes, respectively. Finally, spatially resolved simulations at 12~mT [Fig.~\ref{fig:fig3}(g)] show a uniform phase within each layer, with no internal phase variation.

The opening of a gap at the degeneracy field is the signature of a strong coupling between the acoustic and optical modes. The results obtained by simulations support the experimental results (Fig.~\ref{fig:fig2}), corroborating a strong coupling between the modes. We argue that this strong coupling between acoustic and optical magnons is facilitated by exchange interaction induced by broken intrinsic symmetry in the studied synthetic ferrimagnet \cite{l_d_landau_quantum_1977}.

The macrospin model and micromagnetic simulations further reveal the dependence of the coupling gap \( g \) and the corresponding field \( H_\mathrm{g} \) on the quadratic and biquadratic exchange interactions. As shown in Figs.~\ref{fig:fig4}(a,b), \( H_\mathrm{g} \) increases linearly with both \( J_\mathrm{q} \) and \( J_\mathrm{bq} \). In contrast, the gap size \( g \) exhibits a sublinear power-law dependence on \( J_\mathrm{q} \), with the scaling behavior governed by the strength of \( J_\mathrm{bq} \). 

The dependence of \( g \) on \( J_\mathrm{q} \) can be modeled as:
\begin{equation}
    g = a\, J_\mathrm{q}^{\,b} + c ,
    \label{eq: 9}
    \tag{9}
\end{equation}
where the fitting parameters \( a \), \( b \), and \( c \) depend on the biquadratic exchange interaction. For \( J_\mathrm{bq} = 2.16~\mathrm{kJ/m^3} \), we obtain a fitted exponent of \( b = 0.79 \) [Fig.~\ref{fig:fig4}(a)]. More generally, the exponent \( b \) increases with increasing \( J_\mathrm{bq} \), starting from \( b = 0.5 \) in the absence of biquadratic exchange (\( J_\mathrm{bq} = 0 \)) and approaching a linear scaling regime (\( b = 1 \)) at large \( J_\mathrm{bq} \), as shown in Fig.~\ref{fig:fig4}(c).

The dependence of the gap \( g \) on the biquadratic exchange interaction \( J_\mathrm{bq} \) is regime dependent. For \( J_\mathrm{q} < 2J_\mathrm{bq} \), where a fully antiferromagnetic alignment is not energetically accessible, \( g \) follows a sublinear power-law dependence on \( J_\mathrm{bq} \), with the scaling exponent determined by \( J_\mathrm{q} \). In contrast, for \( J_\mathrm{q} > 2J_\mathrm{bq} \), where a fully antiferromagnetic state becomes accessible [see Eq.~\eqref{eq.2b}], the gap \( g \) decreases with increasing \( J_\mathrm{bq} \) and deviates from the sublinear power-law behavior.

For \( J_\mathrm{q} = 30~\mathrm{kJ/m^3} \), we obtain an approximately linear dependence of \( g \) on \( J_\mathrm{bq} \) in the regime \( J_\mathrm{q} > 2J_\mathrm{bq} \), with a fitted exponent of \( b = 1.02 \) [Fig.~\ref{fig:fig4}(b)]. More generally, the exponent \( b \) increases with increasing \( J_\mathrm{q} \), starting from \( b = 0.5 \) in the absence of quadratic exchange (\( J_\mathrm{q} = 0 \)) and approaching a linear scaling regime (\( b = 1 \)) at large \( J_\mathrm{q} \), as summarized in Fig.~\ref{fig:fig4}(d). Further details are provided in the SM.

\section{Summary}
In summary, we demonstrated strong coupling between acoustic and optical modes in a synthetic ferrimagnet, CoFe(10~nm)/Ru($x$~nm)/NiFe(10~nm), stabilized by broken intrinsic symmetry in an in-plane geometry. Macrospin model and micromagnetic simulations reveal that both the gap field and gap size depend sensitively on the quadratic and biquadratic exchange interactions, $J_\mathrm{q}$ and $J_\mathrm{bq}$. While $J_\mathrm{q}$ and $J_\mathrm{bq}$ can in principle be extracted from magnetostatics via Eqs.~(\ref{eq.3a}) and (\ref{eq.3b}), this approach lacks precision and fails for $J_\mathrm{q}<2J_\mathrm{bq}$ where $H_{0,\mathrm{afm}}$ does not exist. By contrast, the macrospin model developed here provides a robust route to parameter extraction by fitting directly to spin-dynamics spectra. Moreover, micromagnetic simulations enable clear distinction between acoustic and optical modes and, through phase analysis, reveal hybridization at the degeneracy field. Our micromagnetic framework, incorporating both quadratic and biquadratic interactions, is broadly applicable to unconventional geometries, including artificial spin-ice structures \cite{sultana_ice_2025, dion_ultrastrong_2024}, where the macrospin models break down. Furthermore, our study introduces intrinsic symmetry-breaking as a mechanism to tune magnon-magnon coupling in an in-plane geometry, which offers promising prospects for the development of reconfigurable magnonic devices, such as tunable filters, non-reciprocal components, and coherent magnon-based logic elements. The ability to engineer and control hybridized modes through intrinsic symmetry-breaking can significantly enhance the design flexibility of future on-chip spintronic and magnonic circuits.

\section*{Acknowledgments}

This research was supported by NSF through the University of Delaware Materials Research Science and Engineering Center, DMR-2011824. The authors acknowledge the use of facilities and instrumentation supported by NSF through the University of Delaware Materials Research Science and Engineering Center, DMR-2011824.

The data supporting this study's findings are at [?].


\bibliography{references}

@misc{zhang_perspective_2025,
	title = {Perspective: {Magnon}-magnon coupling in hybrid magnonics},
	url = {https://arxiv.org/abs/2511.21904},
	author = {Zhang, Wei and Xiong, Yuzan and Hu, Jia-Mian and Sklenar, Joseph and Subedi, Mitra Mani and Jungfleisch, M. Benjamin and Bhat, Vinayak S. and Li, Yi and Liu, Luqiao and Wang, Qiuyuan and Luo, Yunqiu Kelly and Bae, Youn Jue and Flebus, Benedetta},
	year = {2025},
	note = {\_eprint: 2511.21904},
}

@article{sultana_ice_2025,
	title = {Ice sculpting: {An} artificial spin ice {Tutorial} on controlling microstate and geometry for magnonics and neuromorphic computing},
	volume = {138},
	issn = {0021-8979},
	shorttitle = {Ice sculpting},
	url = {https://doi.org/10.1063/5.0274799},
	doi = {10.1063/5.0274799},
	number = {6},
	urldate = {2025-12-19},
	journal = {Journal of Applied Physics},
	author = {Sultana, Rawnak and Mondal, Amrit Kumar and Bhat, Vinayak Shantaram and Stenning, Kilian and Li, Yue and Arroo, Daan M. and Vasdev, Aastha and McCarter, Margaret R. and De Long, Lance E. and Hastings, J. Todd and Gartside, Jack C. and Jungfleisch, M. Benjamin},
	month = aug,
	year = {2025},
	pages = {061101},
}

@article{dion_ultrastrong_2024,
	title = {Ultrastrong magnon-magnon coupling and chiral spin-texture control in a dipolar {3D} multilayered artificial spin-vortex ice},
	volume = {15},
	url = {https://www.nature.com/articles/s41467-024-48080-z},
	number = {1},
	urldate = {2025-12-19},
	journal = {Nature communications},
	author = {Dion, Troy and Stenning, Kilian D. and Vanstone, Alex and Holder, Holly H. and Sultana, Rawnak and Alatteili, Ghanem and Martinez, Victoria and Kaffash, Mojtaba Taghipour and Kimura, Takashi and Oulton, Rupert F.},
	year = {2024},
	pages = {4077},
}

@article{gilbert_phenomenological_2004,
	title = {A phenomenological theory of damping in ferromagnetic materials},
	volume = {40},
	issn = {1941-0069},
	url = {https://ieeexplore.ieee.org/document/1353448/authors},
	doi = {10.1109/TMAG.2004.836740},
	number = {6},
	urldate = {2025-12-17},
	journal = {IEEE Transactions on Magnetics},
	author = {Gilbert, T.L.},
	month = nov,
	year = {2004},
	pages = {3443--3449},
}

@article{landau_theory_1935,
	title = {On the theory of the dispersion of magnetic permeability in ferromagnetic bodies},
	volume = {8},
	url = {https://cds.cern.ch/record/437299},
	urldate = {2025-10-31},
	journal = {Phys. Z. Sowjet.},
	author = {Landau, Lev Davidovich and Lifshitz, E},
	year = {1935},
	pages = {153},
}

@article{hossain_probing_2021,
	title = {Probing anisotropy in epitaxial {Fe}/{Pt} bilayers by spin–orbit torque ferromagnetic resonance},
	volume = {119},
	url = {https://doi.org/10.1063%2F5.0071151},
	doi = {10.1063/5.0071151},
	number = {21},
	journal = {Applied Physics Letters},
	author = {Hossain, Mohammad Tomal and Lendinez, Sergi and Scheuer, Laura and Papaioannou, Evangelos Th and Jungfleisch, M. Benjamin},
	month = nov,
	year = {2021},
	pages = {212407},
}

@article{subedi_engineering_2025,
	title = {Engineering antiferromagnetic magnon bands through interlayer spin pumping},
	volume = {23},
	issn = {2331-7019},
	url = {http://dx.doi.org/10.1103/PhysRevApplied.23.L031003},
	doi = {10.1103/physrevapplied.23.l031003},
	number = {3},
	journal = {Physical Review Applied},
	author = {Subedi, M.M. and Deng, K. and Xiong, Y. and Mongeon, J. and Hossain, M.T. and Meisenheimer, P.B. and Zhou, E.T. and Heron, J.T. and Jungfleisch, M.B. and Zhang, W. and Flebus, B. and Sklenar, J.},
	month = mar,
	year = {2025},
}

@article{xiao_magnon_2022,
	title = {Magnon mode transition in synthetic antiferromagnets induced by perpendicular magnetic anisotropy},
	volume = {131},
	issn = {0021-8979},
	url = {https://doi.org/10.1063/5.0079266},
	doi = {10.1063/5.0079266},
	number = {9},
	urldate = {2025-09-09},
	journal = {Journal of Applied Physics},
	author = {Xiao, Xiao and Chen, Zhengdong and Dai, Changting and Ma, Fusheng},
	month = mar,
	year = {2022},
	pages = {093905},
}

@article{neumann_uber_1929,
	title = {Uber das {Verhalten} von {Eigenwerten} bei adiabatischen {Prozessen}},
	volume = {30},
	journal = {Physikalische Zeitschrift},
	author = {Neumann, John von and Wigner, Eugene Paul},
	year = {1929},
	pages = {467--470},
}

@article{subedi_even-odd-layer-dependent_2024,
	title = {Even-odd-layer-dependent symmetry breaking in synthetic antiferromagnets},
	volume = {36},
	issn = {0953-8984},
	url = {https://dx.doi.org/10.1088/1361-648X/ad5508},
	doi = {10.1088/1361-648X/ad5508},
	language = {en},
	number = {37},
	urldate = {2024-10-04},
	journal = {Journal of Physics: Condensed Matter},
	author = {Subedi, M. M. and Deng, K. and Flebus, B. and Sklenar, J.},
	month = jun,
	year = {2024},
	pages = {375802},
}

@article{macneill_gigahertz_2019,
	title = {Gigahertz {Frequency} {Antiferromagnetic} {Resonance} and {Strong} {Magnon}-{Magnon} {Coupling} in the {Layered} {Crystal} \{{\textbackslash}textbackslashmathrm\{{CrCl}\}\}\_\{3\}},
	volume = {123},
	url = {https://link.aps.org/doi/10.1103/PhysRevLett.123.047204},
	doi = {10.1103/PhysRevLett.123.047204},
	number = {4},
	urldate = {2024-10-04},
	journal = {Physical Review Letters},
	author = {MacNeill, David and Hou, Justin T. and Klein, Dahlia R. and Zhang, Pengxiang and Jarillo-Herrero, Pablo and Liu, Luqiao},
	month = jul,
	year = {2019},
	pages = {047204},
}

@article{wang_ultrastrong_2024,
	title = {Ultrastrong to nearly deep-strong magnon-magnon coupling with a high degree of freedom in synthetic antiferromagnets},
	volume = {15},
	copyright = {2024 The Author(s)},
	issn = {2041-1723},
	url = {https://www.nature.com/articles/s41467-024-46474-7},
	doi = {10.1038/s41467-024-46474-7},
	language = {en},
	number = {1},
	urldate = {2024-09-13},
	journal = {Nature Communications},
	author = {Wang, Yuqiang and Zhang, Yu and Li, Chaozhong and Wei, Jinwu and He, Bin and Xu, Hongjun and Xia, Jihao and Luo, Xuming and Li, Jiahui and Dong, Jing and He, Wenqing and Yan, Zhengren and Yang, Wenlong and Ma, Fusheng and Chai, Guozhi and Yan, Peng and Wan, Caihua and Han, Xiufeng and Yu, Guoqiang},
	month = mar,
	year = {2024},
	pages = {2077},
}

@article{demokritov_biquadratic_1998,
	title = {Biquadratic interlayer coupling in layered magnetic systems},
	volume = {31},
	issn = {0022-3727},
	url = {https://dx.doi.org/10.1088/0022-3727/31/8/003},
	doi = {10.1088/0022-3727/31/8/003},
	language = {en},
	number = {8},
	urldate = {2024-09-13},
	journal = {Journal of Physics D: Applied Physics},
	author = {Demokritov, S. O.},
	month = apr,
	year = {1998},
	pages = {925},
}

@article{shiota_tunable_2020,
	title = {Tunable {Magnon}-{Magnon} {Coupling} {Mediated} by {Dynamic} {Dipolar} {Interaction} in {Synthetic} {Antiferromagnets}},
	volume = {125},
	url = {https://link.aps.org/doi/10.1103/PhysRevLett.125.017203},
	doi = {10.1103/PhysRevLett.125.017203},
	number = {1},
	urldate = {2023-05-11},
	journal = {Physical Review Letters},
	author = {Shiota, Yoichi and Taniguchi, Tomohiro and Ishibashi, Mio and Moriyama, Takahiro and Ono, Teruo},
	month = jul,
	year = {2020},
	pages = {017203},
}

@article{sud_tunable_2020,
	title = {Tunable magnon-magnon coupling in synthetic antiferromagnets},
	volume = {102},
	url = {https://link.aps.org/doi/10.1103/PhysRevB.102.100403},
	doi = {10.1103/PhysRevB.102.100403},
	number = {10},
	urldate = {2023-01-17},
	journal = {Physical Review B},
	author = {Sud, A. and Zollitsch, C. W. and Kamimaki, A. and Dion, T. and Khan, S. and Iihama, S. and Mizukami, S. and Kurebayashi, H.},
	month = sep,
	year = {2020},
	pages = {100403},
}

@article{liu_spin-torque_2011,
	title = {Spin-{Torque} {Ferromagnetic} {Resonance} {Induced} by the {Spin} {Hall} {Effect}},
	volume = {106},
	url = {https://link.aps.org/doi/10.1103/PhysRevLett.106.036601},
	doi = {10.1103/PhysRevLett.106.036601},
	number = {3},
	urldate = {2023-01-25},
	journal = {Physical Review Letters},
	author = {Liu, Luqiao and Moriyama, Takahiro and Ralph, D. C. and Buhrman, R. A.},
	month = jan,
	year = {2011},
	pages = {036601},
}

@article{sklenar_self-hybridization_2021,
	title = {Self-{Hybridization} and {Tunable} {Magnon}-{Magnon} {Coupling} in van der {Waals} {Synthetic} {Magnets}},
	volume = {15},
	issn = {2331-7019},
	url = {https://link.aps.org/doi/10.1103/PhysRevApplied.15.044008},
	doi = {10.1103/PhysRevApplied.15.044008},
	language = {en},
	number = {4},
	urldate = {2023-01-17},
	journal = {Physical Review Applied},
	author = {Sklenar, Joseph and Zhang, Wei},
	month = apr,
	year = {2021},
	pages = {044008},
}

@article{rongione_emission_2023,
	title = {Emission of coherent {THz} magnons in an antiferromagnetic insulator triggered by ultrafast spin–phonon interactions},
	volume = {14},
	copyright = {2023 The Author(s)},
	issn = {2041-1723},
	url = {https://www.nature.com/articles/s41467-023-37509-6},
	doi = {10.1038/s41467-023-37509-6},
	language = {en},
	number = {1},
	urldate = {2023-05-02},
	journal = {Nature Communications},
	author = {Rongione, E. and Gueckstock, O. and Mattern, M. and Gomonay, O. and Meer, H. and Schmitt, C. and Ramos, R. and Kikkawa, T. and Mičica, M. and Saitoh, E. and Sinova, J. and Jaffrès, H. and Mangeney, J. and Goennenwein, S. T. B. and Geprägs, S. and Kampfrath, T. and Kläui, M. and Bargheer, M. and Seifert, T. S. and Dhillon, S. and Lebrun, R.},
	month = mar,
	year = {2023},
	pages = {1818},
}

@article{chumak_advances_2022,
	title = {Advances in {Magnetics} {Roadmap} on {Spin}-{Wave} {Computing}},
	volume = {58},
	issn = {1941-0069},
	url = {https://ieeexplore.ieee.org/abstract/document/9706176},
	doi = {10.1109/TMAG.2022.3149664},
	number = {6},
	urldate = {2024-09-03},
	journal = {IEEE Transactions on Magnetics},
	author = {Chumak, A. V. and Kabos, P. and Wu, M. and Abert, C. and Adelmann, C. and Adeyeye, A. O. and Åkerman, J. and Aliev, F. G. and Anane, A. and Awad, A. and Back, C. H. and Barman, A. and Bauer, G. E. W. and Becherer, M. and Beginin, E. N. and Bittencourt, V. A. S. V. and Blanter, Y. M. and Bortolotti, P. and Boventer, I. and Bozhko, D. A. and Bunyaev, S. A. and Carmiggelt, J. J. and Cheenikundil, R. R. and Ciubotaru, F. and Cotofana, S. and Csaba, G. and Dobrovolskiy, O. V. and Dubs, C. and Elyasi, M. and Fripp, K. G. and Fulara, H. and Golovchanskiy, I. A. and Gonzalez-Ballestero, C. and Graczyk, P. and Grundler, D. and Gruszecki, P. and Gubbiotti, G. and Guslienko, K. and Haldar, A. and Hamdioui, S. and Hertel, R. and Hillebrands, B. and Hioki, T. and Houshang, A. and Hu, C.-M. and Huebl, H. and Huth, M. and Iacocca, E. and Jungfleisch, M. B. and Kakazei, G. N. and Khitun, A. and Khymyn, R. and Kikkawa, T. and Kläui, M. and Klein, O. and Kłos, J. W. and Knauer, S. and Koraltan, S. and Kostylev, M. and Krawczyk, M. and Krivorotov, I. N. and Kruglyak, V. V. and Lachance-Quirion, D. and Ladak, S. and Lebrun, R. and Li, Y. and Lindner, M. and Macêdo, R. and Mayr, S. and Melkov, G. A. and Mieszczak, S. and Nakamura, Y. and Nembach, H. T. and Nikitin, A. A. and Nikitov, S. A. and Novosad, V. and Otálora, J. A. and Otani, Y. and Papp, A. and Pigeau, B. and Pirro, P. and Porod, W. and Porrati, F. and Qin, H. and Rana, B. and Reimann, T. and Riente, F. and Romero-Isart, O. and Ross, A. and Sadovnikov, A. V. and Safin, A. R. and Saitoh, E. and Schmidt, G. and Schultheiss, H. and Schultheiss, K. and Serga, A. A. and Sharma, S. and Shaw, J. M. and Suess, D. and Surzhenko, O. and Szulc, K. and Taniguchi, T. and Urbánek, M. and Usami, K. and Ustinov, A. B. and van der Sar, T. and van Dijken, S. and Vasyuchka, V. I. and Verba, R. and Kusminskiy, S. Viola and Wang, Q. and Weides, M. and Weiler, M. and Wintz, S. and Wolski, S. P. and Zhang, X.},
	month = jun,
	year = {2022},
	pages = {1--72},
}

@article{wang_nanoscale_2024,
	title = {Nanoscale magnonic networks},
	volume = {21},
	url = {https://link.aps.org/doi/10.1103/PhysRevApplied.21.040503},
	doi = {10.1103/PhysRevApplied.21.040503},
	number = {4},
	urldate = {2024-09-03},
	journal = {Physical Review Applied},
	author = {Wang, Qi and Csaba, Gyorgy and Verba, Roman and Chumak, Andrii V. and Pirro, Philipp},
	month = apr,
	year = {2024},
	pages = {040503},
}

@article{chumak_fundamentals_2019,
	title = {Fundamentals of magnon-based computing},
	volume = {1901.08934},
	url = {http://arxiv.org/abs/1901.08934},
	urldate = {2023-09-22},
	journal = {Arxiv},
	author = {Chumak, A. V.},
	month = jan,
	year = {2019},
}

@article{jungfleisch_high-frequency_2017,
	title = {High-{Frequency} {Dynamics} {Modulated} by {Collective} {Magnetization} {Reversal} in {Artificial} {Spin} {Ice}},
	volume = {8},
	url = {https://link.aps.org/doi/10.1103/PhysRevApplied.8.064026},
	doi = {10.1103/PhysRevApplied.8.064026},
	number = {6},
	urldate = {2024-08-30},
	journal = {Physical Review Applied},
	author = {Jungfleisch, Matthias B. and Sklenar, Joseph and Ding, Junjia and Park, Jungsik and Pearson, John E. and Novosad, Valentine and Schiffer, Peter and Hoffmann, Axel},
	month = dec,
	year = {2017},
	pages = {064026},
}

@article{wu_principles_2021,
	title = {Principles of spintronic {THz} emitters},
	volume = {130},
	issn = {0021-8979},
	url = {https://doi.org/10.1063/5.0057536},
	doi = {10.1063/5.0057536},
	number = {9},
	urldate = {2024-08-30},
	journal = {Journal of Applied Physics},
	author = {Wu, Weipeng and Yaw Ameyaw, Charles and Doty, Matthew F. and Jungfleisch, M. Benjamin},
	month = sep,
	year = {2021},
	pages = {091101},
}

@article{gong_electrically_2018,
	title = {Electrically induced {2D} half-metallic antiferromagnets and spin field effect transistors},
	volume = {115},
	url = {https://www.pnas.org/doi/full/10.1073/pnas.1715465115},
	doi = {10.1073/pnas.1715465115},
	number = {34},
	urldate = {2024-08-22},
	journal = {Proceedings of the National Academy of Sciences},
	author = {Gong, Shi-Jing and Gong, Cheng and Sun, Yu-Yun and Tong, Wen-Yi and Duan, Chun-Gang and Chu, Jun-Hao and Zhang, Xiang},
	month = aug,
	year = {2018},
	pages = {8511--8516},
}

@book{l_d_landau_quantum_1977,
	series = {Course of {Theoretical} {Physics}},
	title = {Quantum {Mechanics} - 3rd {Edition}},
	volume = {Volume 3},
	isbn = {978-1-4831-4912-7},
	author = {{L D Landau} and Lifshitz, E. M.},
	month = may,
	year = {1977},
}

@article{chen_tuning_2020,
	title = {Tuning {Optical} {Mode} of {Ferromagnetic} {Resonance} in {Exchange}-{Coupled} {CoFe}/{Ta}/({NiFe})1–{xCuₓ} {Trilayers}},
	volume = {56},
	issn = {1941-0069},
	url = {https://ieeexplore.ieee.org/document/9119459},
	doi = {10.1109/TMAG.2020.3002974},
	number = {8},
	urldate = {2024-08-21},
	journal = {IEEE Transactions on Magnetics},
	author = {Chen, Hang and Chen, Yunpeng and Wang, Tao and Xie, Yunsong and Franco, Andres F. and Xiao, John Q.},
	month = aug,
	year = {2020},
	pages = {1--6},
}

@article{meier_quantum_2003,
	title = {Quantum computing with antiferromagnetic spin clusters},
	volume = {68},
	url = {https://link.aps.org/doi/10.1103/PhysRevB.68.134417},
	doi = {10.1103/PhysRevB.68.134417},
	number = {13},
	urldate = {2024-08-21},
	journal = {Physical Review B},
	author = {Meier, Florian and Levy, Jeremy and Loss, Daniel},
	month = oct,
	year = {2003},
	pages = {134417},
}

@article{sud_magnon-magnon_2023,
	title = {Magnon-magnon coupling in synthetic ferrimagnets},
	volume = {108},
	url = {https://link.aps.org/doi/10.1103/PhysRevB.108.104407},
	doi = {10.1103/PhysRevB.108.104407},
	number = {10},
	urldate = {2024-08-21},
	journal = {Physical Review B},
	author = {Sud, A. and Yamamoto, K. and Suzuki, K. Z. and Mizukami, S. and Kurebayashi, H.},
	month = sep,
	year = {2023},
	pages = {104407},
}

@article{mahmoud_introduction_2020,
	title = {Introduction to spin wave computing},
	volume = {128},
	issn = {0021-8979},
	url = {https://doi.org/10.1063/5.0019328},
	doi = {10.1063/5.0019328},
	number = {16},
	urldate = {2024-08-21},
	journal = {Journal of Applied Physics},
	author = {Mahmoud, Abdulqader and Ciubotaru, Florin and Vanderveken, Frederic and Chumak, Andrii V. and Hamdioui, Said and Adelmann, Christoph and Cotofana, Sorin},
	month = oct,
	year = {2020},
	pages = {161101},
}

@article{waring_zero-field_2020,
	title = {Zero-field {Optic} {Mode} {Beyond} 20 {GHz} in a {Synthetic} {Antiferromagnet}},
	volume = {13},
	url = {https://link.aps.org/doi/10.1103/PhysRevApplied.13.034035},
	doi = {10.1103/PhysRevApplied.13.034035},
	number = {3},
	urldate = {2023-07-05},
	journal = {Physical Review Applied},
	author = {Waring, H. J. and Johansson, N. A. B. and Vera-Marun, I. J. and Thomson, T.},
	month = mar,
	year = {2020},
	pages = {034035},
}

@article{bowlan_using_2018,
	title = {Using ultrashort terahertz pulses to directly probe spin dynamics in insulating antiferromagnets},
	volume = {51},
	issn = {0022-3727},
	url = {https://dx.doi.org/10.1088/1361-6463/aab8da},
	doi = {10.1088/1361-6463/aab8da},
	language = {en},
	number = {19},
	urldate = {2023-09-22},
	journal = {Journal of Physics D: Applied Physics},
	author = {Bowlan, P. and Trugman, S. A. and Yarotski, D. A. and Taylor, A. J. and Prasankumar, R. P.},
	month = apr,
	year = {2018},
	pages = {194003},
}

@article{chen_unidirectional_2021,
	title = {Unidirectional spin-wave propagation and devices},
	volume = {55},
	issn = {0022-3727},
	url = {https://dx.doi.org/10.1088/1361-6463/ac31f4},
	doi = {10.1088/1361-6463/ac31f4},
	language = {en},
	number = {12},
	urldate = {2023-09-22},
	journal = {Journal of Physics D: Applied Physics},
	author = {Chen, Jilei and Yu, Haiming and Gubbiotti, Gianluca},
	month = nov,
	year = {2021},
	pages = {123001},
}

@article{olejnik_terahertz_2018,
	title = {Terahertz electrical writing speed in an antiferromagnetic memory},
	volume = {4},
	url = {https://www.science.org/doi/10.1126/sciadv.aar3566},
	doi = {10.1126/sciadv.aar3566},
	number = {3},
	urldate = {2023-09-22},
	journal = {Science Advances},
	author = {Olejník, Kamil and Seifert, Tom and Kašpar, Zdeněk and Novák, Vít and Wadley, Peter and Campion, Richard P. and Baumgartner, Manuel and Gambardella, Pietro and Němec, Petr and Wunderlich, Joerg and Sinova, Jairo and Kužel, Petr and Müller, Melanie and Kampfrath, Tobias and Jungwirth, Tomas},
	month = mar,
	year = {2018},
	pages = {eaar3566},
}

@article{lepadatu_synthetic_2017,
	title = {Synthetic ferrimagnet nanowires with very low critical current density for coupled domain wall motion},
	volume = {7},
	copyright = {2017 The Author(s)},
	issn = {2045-2322},
	url = {https://www.nature.com/articles/s41598-017-01748-7},
	doi = {10.1038/s41598-017-01748-7},
	language = {en},
	number = {1},
	urldate = {2023-09-28},
	journal = {Scientific Reports},
	author = {Lepadatu, Serban and Saarikoski, Henri and Beacham, Robert and Benitez, Maria Jose and Moore, Thomas A. and Burnell, Gavin and Sugimoto, Satoshi and Yesudas, Daniel and Wheeler, May C. and Miguel, Jorge and Dhesi, Sarnjeet S. and McGrouther, Damien and McVitie, Stephen and Tatara, Gen and Marrows, Christopher H.},
	month = may,
	year = {2017},
	pages = {1640},
}

@article{patchett_symmetry_2022,
	title = {Symmetry effects on the static and dynamic properties of coupled magnetic oscillators},
	volume = {105},
	url = {https://link.aps.org/doi/10.1103/PhysRevB.105.104436},
	doi = {10.1103/PhysRevB.105.104436},
	number = {10},
	urldate = {2023-09-27},
	journal = {Physical Review B},
	author = {Patchett, J. P. and Drouhin, M. and Liao, J. W. and Soban, Z. and Petit, D. and Haigh, J. and Roy, P. and Wunderlich, J. and Cowburn, R. P. and Ciccarelli, C.},
	month = mar,
	year = {2022},
	pages = {104436},
}

@article{li_symmetry_2021,
	title = {Symmetry breaking induced magnon-magnon coupling in synthetic antiferromagnets},
	volume = {103},
	url = {https://link.aps.org/doi/10.1103/PhysRevB.103.064429},
	doi = {10.1103/PhysRevB.103.064429},
	number = {6},
	urldate = {2024-03-15},
	journal = {Physical Review B},
	author = {Li, Mei and Lu, Jie and He, Wei},
	month = feb,
	year = {2021},
	pages = {064429},
}

@article{grollier_spintronic_2016,
	title = {Spintronic {Nanodevices} for {Bioinspired} {Computing}},
	volume = {104},
	issn = {1558-2256},
	url = {https://ieeexplore.ieee.org/document/7563364},
	doi = {10.1109/JPROC.2016.2597152},
	number = {10},
	urldate = {2024-01-22},
	journal = {Proceedings of the IEEE},
	author = {Grollier, Julie and Querlioz, Damien and Stiles, Mark D.},
	month = oct,
	year = {2016},
	pages = {2024--2039},
}

@article{sankey_spin-transfer-driven_2006,
	title = {Spin-{Transfer}-{Driven} {Ferromagnetic} {Resonance} of {Individual} {Nanomagnets}},
	volume = {96},
	url = {https://link.aps.org/doi/10.1103/PhysRevLett.96.227601},
	doi = {10.1103/PhysRevLett.96.227601},
	number = {22},
	urldate = {2024-08-16},
	journal = {Physical Review Letters},
	author = {Sankey, J. C. and Braganca, P. M. and Garcia, A. G. F. and Krivorotov, I. N. and Buhrman, R. A. and Ralph, D. C.},
	month = jun,
	year = {2006},
	pages = {227601},
}

@article{hrabec_spin-orbit_2016,
	title = {Spin-orbit interaction enhancement in permalloy thin films by {Pt} doping},
	volume = {93},
	url = {https://link.aps.org/doi/10.1103/PhysRevB.93.014432},
	doi = {10.1103/PhysRevB.93.014432},
	number = {1},
	urldate = {2023-07-06},
	journal = {Physical Review B},
	author = {Hrabec, A. and Gonçalves, F. J. T. and Spencer, C. S. and Arenholz, E. and N'Diaye, A. T. and Stamps, R. L. and Marrows, Christopher H.},
	month = jan,
	year = {2016},
	pages = {014432},
}

@article{gui_realization_2007,
	title = {Realization of a {Room}-{Temperature} {Spin} {Dynamo}: {The} {Spin} {Rectification} {Effect}},
	volume = {98},
	shorttitle = {Realization of a {Room}-{Temperature} {Spin} {Dynamo}},
	url = {https://link.aps.org/doi/10.1103/PhysRevLett.98.107602},
	doi = {10.1103/PhysRevLett.98.107602},
	number = {10},
	urldate = {2024-03-28},
	journal = {Physical Review Letters},
	author = {Gui, Y. S. and Mecking, N. and Zhou, X. and Williams, Gwyn and Hu, C.-M.},
	month = mar,
	year = {2007},
	pages = {107602},
}

@article{kosub_purely_2017,
	title = {Purely antiferromagnetic magnetoelectric random access memory},
	volume = {8},
	copyright = {2017 The Author(s)},
	issn = {2041-1723},
	url = {https://www.nature.com/articles/ncomms13985},
	doi = {10.1038/ncomms13985},
	language = {en},
	number = {1},
	urldate = {2023-09-22},
	journal = {Nature Communications},
	author = {Kosub, Tobias and Kopte, Martin and Hühne, Ruben and Appel, Patrick and Shields, Brendan and Maletinsky, Patrick and Hübner, René and Liedke, Maciej Oskar and Fassbender, Jürgen and Schmidt, Oliver G. and Makarov, Denys},
	month = jan,
	year = {2017},
	pages = {13985},
}

@article{hayashi_observation_2023,
	title = {Observation of mode splitting by magnon–magnon coupling in synthetic antiferromagnets},
	volume = {16},
	issn = {1882-0786},
	url = {https://dx.doi.org/10.35848/1882-0786/acd5a6},
	doi = {10.35848/1882-0786/acd5a6},
	language = {en},
	number = {5},
	urldate = {2023-09-22},
	journal = {Applied Physics Express},
	author = {Hayashi, Daiju and Shiota, Yoichi and Ishibashi, Mio and Hisatomi, Ryusuke and Moriyama, Takahiro and Ono, Teruo},
	month = may,
	year = {2023},
	pages = {053004},
}

@article{urazhdin_nanomagnonic_2014,
	title = {Nanomagnonic devices based on the spin-transfer torque},
	volume = {9},
	copyright = {2014 Springer Nature Limited},
	issn = {1748-3395},
	url = {https://www.nature.com/articles/nnano.2014.88},
	doi = {10.1038/nnano.2014.88},
	language = {en},
	number = {7},
	urldate = {2023-09-21},
	journal = {Nature Nanotechnology},
	author = {Urazhdin, S. and Demidov, V. E. and Ulrichs, H. and Kendziorczyk, T. and Kuhn, T. and Leuthold, J. and Wilde, G. and Demokritov, S. O.},
	month = jul,
	year = {2014},
	pages = {509--513},
}

@article{kiselev_microwave_2003,
	title = {Microwave oscillations of a nanomagnet driven by a spin-polarized current},
	volume = {425},
	copyright = {2003 Macmillan Magazines Ltd.},
	issn = {1476-4687},
	url = {https://www.nature.com/articles/nature01967},
	doi = {10.1038/nature01967},
	language = {en},
	number = {6956},
	urldate = {2024-01-22},
	journal = {Nature},
	author = {Kiselev, S. I. and Sankey, J. C. and Krivorotov, I. N. and Emley, N. C. and Schoelkopf, R. J. and Buhrman, R. A. and Ralph, D. C.},
	month = sep,
	year = {2003},
	pages = {380--383},
}

@article{weber_gilbert_2019,
	title = {Gilbert damping of {CoFe}-alloys},
	volume = {52},
	issn = {0022-3727},
	url = {https://dx.doi.org/10.1088/1361-6463/ab2096},
	doi = {10.1088/1361-6463/ab2096},
	language = {en},
	number = {32},
	urldate = {2024-01-15},
	journal = {Journal of Physics D: Applied Physics},
	author = {Weber, Ramon and Han, Dong-Soo and Boventer, Isabella and Jaiswal, Samridh and Lebrun, Romain and Jakob, Gerhard and Kläui, Mathias},
	month = jun,
	year = {2019},
	pages = {325001},
}

@article{baibich_giant_1988,
	title = {Giant {Magnetoresistance} of (001){Fe}/(001){Cr} {Magnetic} {Superlattices}},
	volume = {61},
	url = {https://link.aps.org/doi/10.1103/PhysRevLett.61.2472},
	doi = {10.1103/PhysRevLett.61.2472},
	number = {21},
	urldate = {2024-01-22},
	journal = {Physical Review Letters},
	author = {Baibich, M. N. and Broto, J. M. and Fert, A. and Van Dau, F. Nguyen and Petroff, F. and Etienne, P. and Creuzet, G. and Friederich, A. and Chazelas, J.},
	month = nov,
	year = {1988},
	pages = {2472--2475},
}

@article{binasch_enhanced_1989,
	title = {Enhanced magnetoresistance in layered magnetic structures with antiferromagnetic interlayer exchange},
	volume = {39},
	url = {https://link.aps.org/doi/10.1103/PhysRevB.39.4828},
	doi = {10.1103/PhysRevB.39.4828},
	number = {7},
	urldate = {2024-01-22},
	journal = {Physical Review B},
	author = {Binasch, G. and Grünberg, P. and Saurenbach, F. and Zinn, W.},
	month = mar,
	year = {1989},
	pages = {4828--4830},
}

@article{olejnik_antiferromagnetic_2017,
	title = {Antiferromagnetic {CuMnAs} multi-level memory cell with microelectronic compatibility},
	volume = {8},
	issn = {2041-1723},
	url = {https://www.ncbi.nlm.nih.gov/pmc/articles/PMC5454531/},
	doi = {10.1038/ncomms15434},
	urldate = {2023-09-22},
	journal = {Nature Communications},
	author = {Olejník, K. and Schuler, V. and Marti, X. and Novák, V. and Kašpar, Z. and Wadley, P. and Campion, R. P. and Edmonds, K. W. and Gallagher, B. L. and Garces, J. and Baumgartner, M. and Gambardella, P. and Jungwirth, T.},
	month = may,
	year = {2017},
	pages = {15434},
}

@article{harder_analysis_2011,
	title = {Analysis of the line shape of electrically detected ferromagnetic resonance},
	volume = {84},
	url = {https://link.aps.org/doi/10.1103/PhysRevB.84.054423},
	doi = {10.1103/PhysRevB.84.054423},
	number = {5},
	urldate = {2023-07-05},
	journal = {Physical Review B},
	author = {Harder, M. and Cao, Z. X. and Gui, Y. S. and Fan, X. L. and Hu, C.-M.},
	month = aug,
	year = {2011},
	pages = {054423},
}

@article{haldar_reconfigurable_2016,
	title = {A reconfigurable waveguide for energy-efficient transmission and local manipulation of information in a nanomagnetic device},
	volume = {11},
	copyright = {2016 Springer Nature Limited},
	issn = {1748-3395},
	url = {https://www.nature.com/articles/nnano.2015.332},
	doi = {10.1038/nnano.2015.332},
	language = {en},
	number = {5},
	urldate = {2023-09-21},
	journal = {Nature Nanotechnology},
	author = {Haldar, Arabinda and Kumar, Dheeraj and Adeyeye, Adekunle Olusola},
	month = may,
	year = {2016},
	pages = {437--443},
}

@article{franco_multi-state_2018,
	title = {A multi-state synthetic ferrimagnet with controllable switching near room temperature},
	volume = {51},
	issn = {0022-3727},
	url = {https://dx.doi.org/10.1088/1361-6463/aabed0},
	doi = {10.1088/1361-6463/aabed0},
	language = {en},
	number = {22},
	urldate = {2023-09-28},
	journal = {Journal of Physics D: Applied Physics},
	author = {Franco, A. F. and Landeros, P.},
	month = may,
	year = {2018},
	pages = {225003},
}

@article{tulapurkar_spin-torque_2005,
	title = {Spin-torque diode effect in magnetic tunnel junctions},
	volume = {438},
	copyright = {2005 Springer Nature Limited},
	issn = {1476-4687},
	url = {https://www.nature.com/articles/nature04207},
	doi = {10.1038/nature04207},
	language = {en},
	number = {7066},
	urldate = {2024-08-16},
	journal = {Nature},
	author = {Tulapurkar, A. A. and Suzuki, Y. and Fukushima, A. and Kubota, H. and Maehara, H. and Tsunekawa, K. and Djayaprawira, D. D. and Watanabe, N. and Yuasa, S.},
	month = nov,
	year = {2005},
	pages = {339--342},
}

@article{xie_engineering_2023,
	title = {Engineering {Spin} {Configurations} of {Synthetic} {Antiferromagnet} by {Controlling} {Long}-{Range} {Oscillatory} {Interlayer} {Coupling} and {Neighboring} {Ferrimagnetic} {Coupling}},
	volume = {35},
	copyright = {© 2022 Wiley-VCH GmbH},
	issn = {1521-4095},
	doi = {10.1002/adma.202208275},
	language = {en},
	number = {2},
	urldate = {2023-09-28},
	journal = {Advanced Materials},
	author = {Xie, Xuejie and Wang, Xiujuan and Wang, Wei and Zhao, Xiaonan and Bai, Lihui and Chen, Yanxue and Tian, Yufeng and Yan, Shishen},
	year = {2023},
	pages = {2208275},
}

@article{khitun_magnonic_2010,
	title = {Magnonic logic circuits},
	volume = {43},
	issn = {0022-3727},
	url = {https://dx.doi.org/10.1088/0022-3727/43/26/264005},
	doi = {10.1088/0022-3727/43/26/264005},
	language = {en},
	number = {26},
	urldate = {2024-01-22},
	journal = {Journal of Physics D: Applied Physics},
	author = {Khitun, Alexander and Bao, Mingqiang and Wang, Kang L.},
	month = jun,
	year = {2010},
	pages = {264005},
}

@article{luo_implementation_2023,
	title = {Implementation of a full {Wheatstone}-bridge {GMR} sensor by utilizing spin–orbit torque induced magnetization switching in synthetic antiferromagnetic layer},
	volume = {133},
	issn = {0021-8979},
	url = {https://doi.org/10.1063/5.0137559},
	doi = {10.1063/5.0137559},
	number = {15},
	urldate = {2024-01-22},
	journal = {Journal of Applied Physics},
	author = {Luo, Keliu and Guo, Yonghai and Li, Wangda and Zhang, Bo and Wang, Bo and Cao, Jiangwei},
	month = apr,
	year = {2023},
	pages = {153902},
}

@article{inoue_theory_1994,
	title = {A theory of biquadratic exchange coupling in magnetic multilayers},
	volume = {136},
	issn = {0304-8853},
	url = {https://www.sciencedirect.com/science/article/pii/0304885394003378},
	doi = {10.1016/0304-8853(94)00337-8},
	number = {3},
	urldate = {2023-10-13},
	journal = {Journal of Magnetism and Magnetic Materials},
	author = {Inoue, J.},
	month = sep,
	year = {1994},
	pages = {233--237},
}

@article{jenkins_current_2014,
	title = {Current driven magnetization dynamics of a self-polarised synthetic ferrimagnet},
	volume = {115},
	issn = {0021-8979},
	url = {https://doi.org/10.1063/1.4866871},
	doi = {10.1063/1.4866871},
	number = {8},
	urldate = {2023-09-28},
	journal = {Journal of Applied Physics},
	author = {Jenkins, A. S. and Lacoste, B. and Geranton, G. and Gusakova, D. and Dieny, B. and Ebels, U. and Buda-Prejbeanu, L. D.},
	month = feb,
	year = {2014},
	pages = {083911},
}

@article{dai_strong_2021,
	title = {Strong magnon–magnon coupling in synthetic antiferromagnets},
	volume = {118},
	issn = {0003-6951},
	url = {https://doi.org/10.1063/5.0041431},
	doi = {10.1063/5.0041431},
	number = {11},
	urldate = {2023-09-22},
	journal = {Applied Physics Letters},
	author = {Dai, Changting and Ma, Fusheng},
	month = mar,
	year = {2021},
	pages = {112405},
}

@article{mashkovich_terahertz_2019,
	title = {Terahertz {Optomagnetism}: {Nonlinear} {THz} {Excitation} of {GHz} {Spin} {Waves} in {Antiferromagnetic} {FeBO} 3},
	volume = {123},
	issn = {0031-9007, 1079-7114},
	shorttitle = {Terahertz {Optomagnetism}},
	url = {https://link.aps.org/doi/10.1103/PhysRevLett.123.157202},
	doi = {10.1103/PhysRevLett.123.157202},
	language = {en},
	number = {15},
	urldate = {2023-09-22},
	journal = {Physical Review Letters},
	author = {Mashkovich, E. A. and Grishunin, K. A. and Mikhaylovskiy, R. V. and Zvezdin, A. K. and Pisarev, R. V. and Strugatsky, M. B. and Christianen, P. C. M. and Rasing, Th. and Kimel, A. V.},
	month = oct,
	year = {2019},
	pages = {157202},
}

@article{lee_terahertz_2021,
	title = {Terahertz spectroscopy of antiferromagnetic resonances in {YFe1}−{xMnxO3} ≤x≤0.4 across a spin reorientation transition},
	volume = {119},
	issn = {0003-6951},
	url = {https://doi.org/10.1063/5.0070952},
	doi = {10.1063/5.0070952},
	number = {19},
	urldate = {2023-09-22},
	journal = {Applied Physics Letters},
	author = {Lee, Howon and Jung, Taek Sun and Shin, Hyun Jun and Oh, Sang Hyup and Sim, Kyung Ik and Ha, Taewoo and Choi, Young Jai and Kim, Jae Hoon},
	month = nov,
	year = {2021},
	pages = {192903},
}

@article{xiong_antiferromagnetic_2022,
	title = {Antiferromagnetic spintronics: {An} overview and outlook},
	volume = {2},
	issn = {2667-3258},
	shorttitle = {Antiferromagnetic spintronics},
	url = {https://www.sciencedirect.com/science/article/pii/S2667325822001443},
	doi = {10.1016/j.fmre.2022.03.016},
	number = {4},
	urldate = {2023-09-22},
	journal = {Fundamental Research},
	author = {Xiong, Danrong and Jiang, Yuhao and Shi, Kewen and Du, Ao and Yao, Yuxuan and Guo, Zongxia and Zhu, Daoqian and Cao, Kaihua and Peng, Shouzhong and Cai, Wenlong and Zhu, Dapeng and Zhao, Weisheng},
	month = jul,
	year = {2022},
	pages = {522--534},
}

@article{vansteenkiste_design_2014,
	title = {The design and verification of {MuMax3}},
	volume = {4},
	issn = {2158-3226},
	url = {https://doi.org/10.1063/1.4899186},
	doi = {10.1063/1.4899186},
	number = {10},
	urldate = {2023-09-11},
	journal = {AIP Advances},
	author = {Vansteenkiste, Arne and Leliaert, Jonathan and Dvornik, Mykola and Helsen, Mathias and Garcia-Sanchez, Felipe and Van Waeyenberge, Bartel},
	month = oct,
	year = {2014},
	pages = {107133},
}

@article{parreiras_effect_2015,
	series = {20th {International} {Conference} on {Magnetism}, {ICM} 2015},
	title = {Effect of {Planar} {Anisotropy} in {Vortex} {Configuration} of {Micro}-scale {Disks}},
	volume = {75},
	issn = {1875-3892},
	url = {https://www.sciencedirect.com/science/article/pii/S1875389215018246},
	doi = {10.1016/j.phpro.2015.12.185},
	language = {en},
	urldate = {2023-07-07},
	journal = {Physics Procedia},
	author = {Parreiras, S. O. and Martins, M. D.},
	month = jan,
	year = {2015},
	pages = {1142--1149},
}

@article{exl_labontes_2014,
	title = {{LaBonte}'s method revisited: {An} effective steepest descent method for micromagnetic energy minimization},
	volume = {115},
	issn = {0021-8979, 1089-7550},
	shorttitle = {{LaBonte}'s method revisited},
	url = {https://pubs.aip.org/jap/article/115/17/17D118/367294/LaBonte-s-method-revisited-An-effective-steepest},
	doi = {10.1063/1.4862839},
	language = {en},
	number = {17},
	urldate = {2023-07-06},
	journal = {Journal of Applied Physics},
	author = {Exl, Lukas and Bance, Simon and Reichel, Franz and Schrefl, Thomas and Peter Stimming, Hans and Mauser, Norbert J.},
	month = may,
	year = {2014},
	pages = {17D118},
}

@book{stancil_spin_2008,
	address = {Berlin},
	title = {Spin waves: theory and applications},
	isbn = {978-0-387-77864-8},
	shorttitle = {Spin waves},
	publisher = {Springer},
	author = {Stancil, Daniel D. and Prabhakar, Anil},
	year = {2008},
}

\end{document}